\documentclass[onecolumn,nofootinbib,floats,superscriptaddress,prd]{revtex4}
\usepackage{graphicx}
\usepackage{amsfonts}
\usepackage{amssymb}
\usepackage{amsbsy}
\usepackage{amsmath}
\usepackage{latexsym}
\usepackage{natbib}
\usepackage{bm}
\usepackage{subfigure}
\usepackage{hyperref}
\usepackage{lscape}
\usepackage{color}
\usepackage{array}
\usepackage{ifthen}
\usepackage{xstring}

\newcommand*{\di}{\partial}

\renewcommand*{\c}{\text{c}}

\renewcommand*{\b}{\text{b}}

\newcommand*{\homo}{{(\text{h})}}
\newcommand*{\particular}{{{(\text{p})}}}

\newcommand*{\Hinf}{H}

\begin{document}

\title{Analytic Description of DGP Perturbations on All Scales}

\author{Sanjeev S.~Seahra}
\affiliation{Department of Mathematics and Statistics, University of
New Brunswick, Fredericton, NB, Canada E3B 5A3}

\author{Wayne Hu}
\affiliation{Department of Astronomy \& Astrophysics, Kavli Institute for Cosmological Physics, and Enrico Fermi Institute, The 
University of
Chicago, Chicago, IL 60637-1433}

\begin{abstract}
We develop analytic solutions for the linear evolution of metric perturbations in the DGP braneworld modified gravity scenario including near-horizon and superhorizon modes where
solutions in the bulk are required.   These solutions apply to both the self-accelerating and
normal branch and
elucidate the nature of coordinate singularities and initial data in the bulk as well as  their effect on perturbation evolution on the brane.      
Even on superhorizon scales, the evolution of metric perturbations is no longer
necessarily scale free due to multiple resonances in the bulk.
Based on these analytic solutions, we devise convenient
fitting functions for the evolution that bridge the various spatial and temporal regimes. Compared with a direct numerical integration of the bulk equations, 
the fits are  accurate at the percent level and are sufficient for current and upcoming  observational tests.
\end{abstract}

\date{\today}

\maketitle

\section{Introduction}

The Dvali-Gabadadze-Porrati (DGP) model modifies General
Relativity on large scales by positing that we live on a 4-dimensional brane in a
5-dimensional Minkowski bulk \cite{Dvali:2000hr}.  On scales larger than the crossover
scale $r_\c$, gravity becomes 5-dimensional and hence one can hope to uncover
extra-dimensional physics by studying the evolution of density perturbations on
scales approaching $r_\c$.
The DGP model has two branches of cosmological solutions \cite{Deffayet:2000uy}.
On the self-accelerating branch, the expansion of the universe accelerates without
a cosmological constant or dark energy.  On the normal branch, brane tension is required
to accelerate the expansion though gravity is still modified on large scales.

To solve for the evolution of perturbations on either branch near the horizon scale and beyond
requires following metric perturbations into the bulk.   On the self-accelerating branch
approximate scaling solutions \cite{Sawicki:2006jj} showed that modifications near the
horizon scale are in significant tension with the data from the cosmic microwave
background \cite{Song:2006jk,Fang:2008kc}.   In this paper, we present the first quantitative 
and thorough comparison of these approximate solutions with direct numeric simulations
\cite{Cardoso:2007xc}.  We confirm that the scaling solutions used in \cite{Song:2006jk,Fang:2008kc} 
are sufficiently accurate to expose tension between self-accelerating DGP and observations.
It is also worthwhile noting that the pure de Sitter phase of the self-accelerating branch suffers from a
perturbative ghost mode (see e.g.~\cite{Koyama:2007za}), which is significant theoretical challenge for the model.

%By comparing these approximations with
%direct numerical integration of the bulk metric equations \cite{Cardoso:2007xc}, 
%we confirm here that these scaling 
%solutions are sufficiently accurate to expose this tension.  Moreover,
%the self-accelerating branch contains a ghost (see e.g.~\cite{Koyama:2007za}).

Numerical integration of the bulk metric equations have also been performed on the ghost-free
normal branch \cite{Cardoso:2007xc}.
In order to facilitate studies of cosmological constraints on the normal branch, it is useful
to have a simple but accurate description of the evolution of perturbations in terms of
the fundamental cosmological parameters and crossover scale.
Scaling approaches on the normal branch have also been studied \cite{Song:2007wd}
but without a proper treatment of initial conditions in the bulk and boundary conditions on
the brane.

 In this paper, we
employ analytic and numerical techniques to devise such a description.
In fact, preliminary versions of results from this work have already been used to set cosmological constraints on the crossover scale 
on both branches of DGP \cite{Lombriser:2009xg}.

We begin in \S \ref{sec:formalism} with a brief review of the equations governing
the background and perturbations on both the self-accelerating and normal branches of
the DGP model.   We outline the numerical methodology for obtaining solutions of
 the perturbation equations
as well as the scaling ansatz that provides insight into their
nature  in \S \ref{sec:methodology}.  In \S \ref{sec:analytic} we combine scaling results with Green's function techniques
to obtain analytic solutions in the matter and de Sitter epochs. We join these
solutions into simple but accurate global descriptions of perturbation evolution via
fitting functions in \S \ref{sec:comparison} and discuss these results in \S \ref{sec:discussion}.

\section{Formalism}
\label{sec:formalism}

We begin in \S \ref{sec:generaleqn} with a review of the DGP field equations and 
bulk geometry.   In \S \ref{sec:background}-\ref{sec:perteqn} we apply the field equations to the background evolution and perturbation evolution respectively.
We discuss the initial conditions and singularities in the bulk in \S \ref{sec:initial}
and the effective equations of motion on the brane in \S \ref{sec:randg}.

\subsection{Field Equations and Bulk Geometry}
\label{sec:generaleqn}

The action of the DGP model is
\begin{equation}
    S =  \frac{1}{2\kappa_5^2} \int\limits_\mathcal{M} d^5 X \sqrt{-g} R^{(5)}
    + \frac{1}{2\kappa_4^2} \int\limits_{\di\mathcal{M}_\b} d^4 x \sqrt{-\gamma} R^{(4)}+
    \int\limits_{\di\mathcal{M}_\b} d^4 x \sqrt{-\gamma} (\mathcal{L}_m
    - \sigma).
\end{equation}
Here, $\sigma$ is the brane tension and $\mathcal{L}_m$ is the
matter Lagrangian.
The
field equations satisfied in the empty bulk are simply
\begin{equation}
    R^{(5)}_{ab} = 0,
    \label{eq:bulkfieldeqn}
\end{equation}
whereas those on the brane are given by 
\cite{shiromizu99,maeda03}
\begin{equation}
\label{eq:4D Einstein equations} G^{(4)}_{\mu\nu} =(2 \kappa_4^2
r_\c)^2 \Pi_{\mu \nu}-\mathcal{E}_{\mu\nu},
\end{equation}
where
\begin{gather}\nonumber
    \Pi_{\mu \nu} =  -\tfrac{1}{4} \tilde{T}_{\mu \alpha}
    \tilde{T}_{\nu}{}^{\alpha} +\tfrac{1}{12} \tilde{T}
    \tilde{T}_{\mu \nu} + \tfrac{1}{24}( 3 \tilde{T}_{\alpha \beta}
    \tilde{T}^{\alpha \beta}- \tilde{T}^2 )
    g_{\mu \nu}, \\
    \tilde{T}_{\mu \nu} =  T_{\mu \nu} - \sigma g_{\alpha\beta}
    - \kappa_4^{-2} G^{(4)}_{\mu \nu},
\end{gather}
and $\mathcal{E}_{\mu \nu}$ is the trace-free projection of the
5-dimensional Weyl tensor.  Here
the crossover scale is defined by
\begin{equation}
    r_\c = \frac{\kappa_5^2}{2\kappa_4^2}.
\end{equation}

The background metric or geometry converts the field equations into the modified
Friedmann equation and imposes the distinction between the two branches
of solutions.
The background bulk metric can be expressed in terms of
Gaussian normal (GN) coordinates $(t,y,\mathbf{x})$
\cite{deffayet00b}
\begin{equation}\label{eq:GN line element}
    ds^2 = -n^2(t,y)\,dt^2 + b^2(t,y) \, d\mathbf{x}^2 + dy^2,
\end{equation}
or null coordinates $(u,v,\mathbf{x})$,
\begin{equation}\label{eq:null line element}
    ds^2 = -du\,dv + r_\c^{-2} v^2 d\mathbf{x}^2.
\end{equation}
For the analytic solutions below, we employ GN coordinates
 whereas for the numerical
characteristic integration (CI) we use null coordinates.

In GN
coordinates the brane is always at $y = 0$, while in the null
coordinates the brane is moving.  The metric functions in the GN line
element are
\begin{equation}
    n(t,y) = 1 + \epsilon\left( H + \frac{\dot{H}}{H} \right)y, \quad
    b(t,y) = a(t)\left( 1 + \epsilon H y \right),
    \label{eq:GNbackgroundmetric}
\end{equation}
where $\epsilon=+1$ on the self-accelerating branch and $\epsilon=-1$ on the normal
branch.  Here $H= \dot a/a$ is the Hubble parameter on the brane.

The coordinate transformation between the two systems is
\begin{equation}\label{eq:coordinate transformation}
    u = \frac{1}{r_\c} \left[ \int_{a_0}^{a(t)}
    \frac{d\tilde{a}}{H^2(\tilde{a}) \tilde{a}^2} - \frac{\epsilon
    y}{Ha} \right] + u_0, \quad v = r_\c b(t,y),
\end{equation}
where $u(a_0)=u_0$ is a constant defining the  origin of
the coordinate system at $a_0$.
Ignoring inflation and taking only matter components on the brane, we can
take $a_0 \rightarrow 0$ and set $u_0=0$.

\begin{figure}
\includegraphics{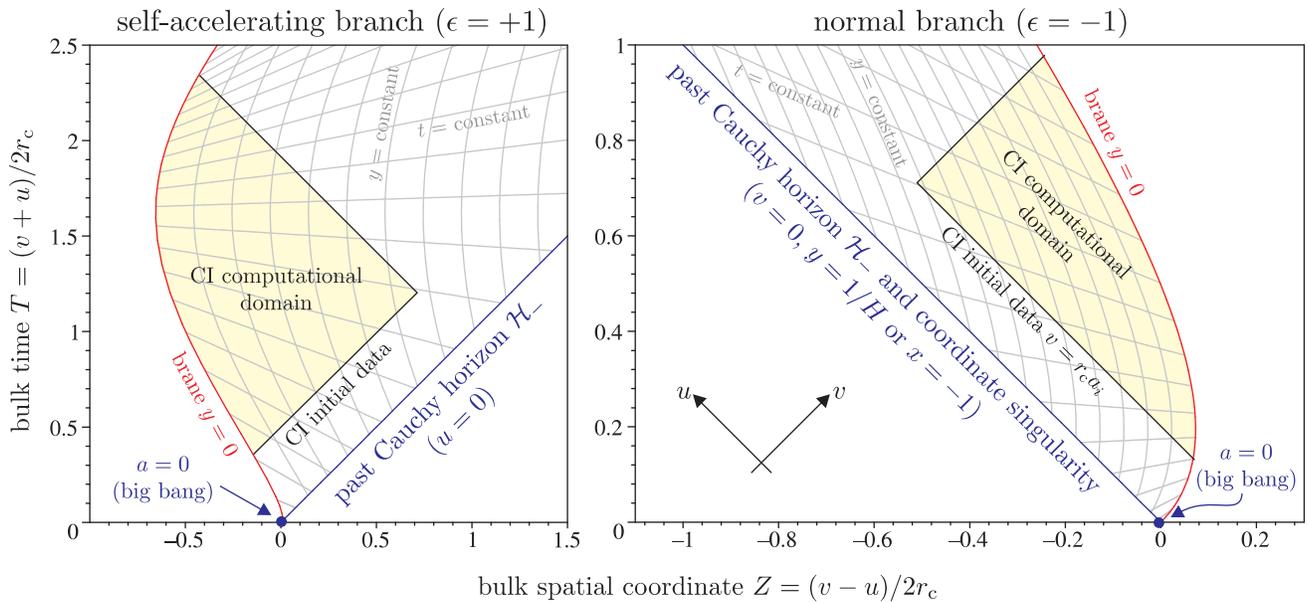}
\caption{Spacetime diagram of a brane trajectory through the bulk
for the self-accelerating and normal branch. The grey timelike
lines are the $y=$ constant traces of the Gaussian normal
coordinate system, while the spacelike grey lines are the $t=$
constant traces. Lines of constant $u$ and $v$ are tilted at
$45^o$ and represent the trajectory of light rays in the bulk. The
brane is at $y = 0$.  In both cases there is a past Cauchy horizon
$\mathcal{H}_-$ in the bulk, and for the normal branch
$\mathcal{H}_-$ is coincident with a coordinate singularity.  In
this plot, we have chosen the constant $u_0$ in Eq.~(\ref{eq:coordinate
transformation}) such that the Cauchy horizon in the
self-accelerating branch is at $u=0$.} \label{fig:bulkdiagram}
\end{figure}

With these assumptions the coordinate transformation fixes the position of the brane in
the null system:
\begin{equation}
    u_\b(t) = \frac{1}{r_\c} \int_0^{a(t)}
    \frac{d\tilde{a}}{H^2(\tilde{a}) \tilde{a}^2}, \quad v_\b(t) =
    r_\c a(t).
    \label{eq:Nullbackgroundmetric}
\end{equation}
Note that both line elements have a coordinate singularity: either
when $b(t,y) = 0$ or $v = 0$.  These singularities are actually at
the same place.  The relationship between the two coordinate
systems is illustrated in Fig.~\ref{fig:bulkdiagram}.  Note that
the $v=0$ singularity is only accessible in the normal branch.

\subsection{Background Evolution}
\label{sec:background}

The field equations in the spatially flat background reduce to
\begin{equation}
    H = \frac{1}{2r_\c} \left[ \epsilon + \sqrt{1 + \frac{4}{3}\kappa_4^2 r_\c^2
    (\rho+\sigma)}\right],
\end{equation}
with the bulk metric determined by $H$ through
Eqs.~({\ref{eq:GNbackgroundmetric}-\ref{eq:Nullbackgroundmetric}).
If we further assume that the energy content of  brane $\rho$ is dominated by non-relativistic matter, this modified Friedmann equation 
can be rewritten
as
\begin{equation}
{H \over H_0} = \sqrt{ \Omega_m a^{-3}
+\Omega_\Lambda + \Omega_{r_\c}} + \epsilon \sqrt{\Omega_{r_\c}},
\end{equation}
where $\Omega_\Lambda= \kappa_4^2 \sigma/3 H_0^2$ is the effective
cosmological constant for the brane tension and 
\begin{equation}
\sqrt{\Omega_{r_\c}} \equiv {1 \over 2 H_0 r_\c} = {\epsilon \over 2}(1-\Omega_m-\Omega_\Lambda),
\end{equation}
defines the crossover scale in terms of the other cosmological parameters.

\subsection{Perturbation Equations}
\label{sec:perteqn}

The perturbed GN line element is written as:
\begin{equation}\label{eq:perturbed metric}
    ds^2 = -n^2(1+2\lambda A)\,dt^2 + 2n \lambda A_y\,dt\,dy +
    b^2(1+2\lambda \mathcal{R}) \, d\mathbf{x}^2 + (1+2\lambda A_{yy}) dy^2,
\end{equation}
in the longitudinal gauge where $\lambda \ll 1$ is an order parameter for bookkeeping
purposes.  These metric perturbations are governed by the master
variable $\Omega$ 
 \cite{mukohyama00}
through the relations
\begin{subequations}\label{eq:potential definitions} \cite{Deffayet:2002fn}
\begin{align}
    A & = -\frac{1}{6b}\left( 2 \frac{\di^2\Omega}{\di y^2} +
    \frac{1}{n^2} \frac{\di^2\Omega}{\di t^2} - \frac{1}{n^3}
    \frac{\di n}{\di t} \frac{\di \Omega}{\di t} - \frac{1}{n}
    \frac{\di n}{\di y} \frac{\di\Omega}{\di y} \right), \\
    A_y & = + \frac{1}{bn} \left( \frac{\di^2\Omega}{\di t \di y} -
    \frac{1}{n} \frac{\di n}{\di y} \frac{\di \Omega}{\di t}
    \right), \\
    A_{yy} & = + \frac{1}{6b}\left( \frac{\di^2\Omega}{\di y^2} +
    \frac{2}{n^2} \frac{\di^2\Omega}{\di t^2} - \frac{2}{n^3}
    \frac{\di n}{\di t} \frac{\di \Omega}{\di t} - \frac{2}{n}
    \frac{\di n}{\di y} \frac{\di\Omega}{\di y} \right), \\
    \mathcal{R} & = + \frac{1}{6b}\left( \frac{\di^2\Omega}{\di y^2}
    -
    \frac{1}{n^2} \frac{\di^2\Omega}{\di t^2} + \frac{1}{n^3}
    \frac{\di n}{\di t} \frac{\di \Omega}{\di t} + \frac{1}{n}
    \frac{\di n}{\di y} \frac{\di\Omega}{\di y} \right).
\end{align}
\end{subequations}
The master variable satisfies the wave, or master, equation
\begin{subequations}\label{eq:wave equations}
\begin{gather}
\label{eq:GN wave eqn}
    -\frac{\di}{\di t} \left( \frac{1}{nb^3} \frac{\di\Omega}{\di t}
    \right)
    + \frac{\di}{\di y} \left( \frac{n}{b^3} \frac{\di\Omega}{\di y}
    \right) - \frac{n}{b^5} k^2 \Omega = 0, \\\label{eq:null wave eqn}
    - \frac{\di^2 \Omega}{\di u \di v} + \frac{3}{2v} \frac{\di\Omega}{\di
    v} - \frac{k^2 r_\c^2}{4v^2} \Omega = 0,
\end{gather}
\end{subequations}
in GN and null coordinates, respectively.  

The master variable also satisfies the boundary condition
\begin{equation}\label{eq:scalar boundary condition}
    (\di_y \Omega)_\b = - \frac{\epsilon \gamma_1}{2H}
    \ddot\Omega_\b +\frac{9\epsilon \gamma_3}{4} \dot\Omega_\b -
    \frac{3(\epsilon\gamma_3 k^2 + \gamma_4 H^2 a^2) }{4H a^2}
    \Omega_\b + \frac{3\epsilon r_\c\kappa_4^2 \rho a^3 \gamma_4 }{2k^2}
    \Delta.
\end{equation}
The boundary value of the master variable $\Omega_\b$ acts as a
source to the brane metric perturbations $\Psi=A(y=0)$ and
$\Phi={\cal R}(y=0)$:
\begin{subequations}\label{eq:Phi and Psi}
\begin{eqnarray}
    \Phi & = & +\frac{\kappa_4^2 \rho a^2 \gamma_1}{2k^2}\Delta +
    \frac{\epsilon \gamma_1}{4ar_\c} \dot\Omega_\b -
    \frac{\epsilon(k^2+3H^2a^2)\gamma_1}{12 Hr_\c a^3} \Omega_\b,  \\
    \Psi & = & -\frac{\kappa_4^2 \rho a^2 \gamma_2}{2k^2}\Delta +
    \frac{\epsilon \gamma_1}{4Hr_\c a } \ddot \Omega_\b -
    \frac{3\epsilon H \gamma_4}{4a} \dot\Omega_\b +
    \frac{\epsilon(k^2 r_\c \gamma_4 + Ha^2 \gamma_2)}{4r_\c
    a^3}\Omega_\b,
\end{eqnarray}
\end{subequations}
where the dimensionless $\gamma$-factors
\begin{subequations}\label{eq:gamma defs}
\begin{align}
    \gamma_1 & = \frac{2\epsilon H r_\c}{2\epsilon H r_\c - 1}, \\
    \gamma_2 & = \frac{2\epsilon r_\c(\dot H - H^2 + 2\epsilon H^3 r_\c)}
    {H(2\epsilon H r_\c -1)^2}, \\
    \gamma_3 & = \frac{4\epsilon r_\c(2\epsilon r_\c \dot H - 3H + 6\epsilon H^2 r_\c)}
    {9 (2\epsilon H r_\c -1)^2}, \\
    \gamma_4 & = \frac{4\epsilon (\epsilon r_\c \dot H - H + 2\epsilon H^2 r_\c)}
    {3 H (2\epsilon H r_\c -1)^2}.
\end{align}
\end{subequations}
With these relations, energy-momentum conservation of the matter on
the brane simplify to yield
\begin{equation}\label{eq:scalar brane equation}
    \ddot \Delta + 2H \dot \Delta - \frac{1}{2} \kappa_4^2 \rho
    \gamma_2 \Delta = - \frac{\epsilon \gamma_4 k^4}{4a^5}
    \Omega_\b.
\end{equation}
Eq. (\ref{eq:wave equations}), (\ref{eq:scalar boundary condition}) and (\ref{eq:scalar brane equation}) represent the complete equations of motion for  the perturbations.

\subsection{Initial Conditions}
\label{sec:initial}

In order to fully specify the dynamics of perturbations, the equations of motion
 must be augmented by initial conditions, not only on the brane but also
in the bulk.  On the brane, the suppression of DGP modifications at
$H r_\c \gg 1$ means that initial conditions can be set as usual with adiabatic
conditions.   In the bulk, it is physically sensible that all bulk perturbations are causally generated by perturbations on the brane; i.e., there should be no sources for $\Omega$ except the brane itself.  This is equivalent to demanding that the bulk master variable vanish on the past Cauchy horizon:
\begin{equation}\label{eq:ICs}
	\Omega(\mathcal{H}_-) = 0.
\end{equation}
While the principal motivation for this assumption is causality, it is interesting to note that in the normal branch this is a necessary condition to avoid a curvature singularity in the bulk.  To see this we substitute Eqs.~(\ref{eq:potential definitions}) and (\ref{eq:GN
wave eqn}) into (\ref{eq:perturbed metric}), and then calculate the full Riemann tensor in the normal branch.
We find that the Riemann tensor tends to diverge on
the past Cauchy horizon $\mathcal{H}_-$ (where $b = 0$ and $v=0$) in the
normal branch unless $\Omega$ is correspondingly suppressed; for example,
\begin{equation}\label{eq:Riemann}
    R_{tyty} = - \frac{\lambda}{3} \frac{k^4 n^2 \Omega }{b^5} +
    \mathcal{O}(\lambda^2).
\end{equation}
We shall see that this issue is related to an irregular singular point
at the horizon at finite $k$ in the master equation and requires
that $\Omega$ be exponentially suppressed near $\mathcal{H}_-$, which is actually a stronger condition than Eq.~(\ref{eq:ICs}).

\subsection{Brane Gradient and Metric}
\label{sec:randg}

The dynamical equations  (\ref{eq:scalar boundary condition}) and (\ref{eq:scalar brane equation}) are
closed on the brane if the relationship between $(\partial_y \Omega)_\b$ and $\Omega_\b$ is known.
  From the perspective
of the observer on the brane, the wave equation (\ref{eq:wave equations})
for the master variable
 in the bulk simply serves to specify the dimensionless gradient
\begin{equation}
R \equiv \left( {\partial_y \Omega \over \epsilon H\Omega } \right)_\b .
\label{eq:gradient}
\end{equation}
We shall see that scaling and Green's function arguments can be used to extract this gradient in various limits.

Ultimately, we are most interested not in the brane gradient $R$, but the metric perturbations themselves. In particular the ratio of metric perturbations
\begin{equation}
g(a,k) \equiv {\Phi+\Psi \over \Phi-\Psi}
\end{equation}
plays a central role in modified gravity models as it vanishes in General Relativity if there
is no anisotropic stress.   Moreover in 
 the parameterized post-Friedmann (PPF) formalism \cite{Hu:2007pj}, it efficiently specifies metric perturbations for all scales.  For the large scale regime $(k/Ha) \ll 1$, in General Relativity it is well known that the curvature perturbation on comoving slices $\zeta$ is conserved.  This conservation law is a consequence of energy-momentum
conservation and applies to DGP as well \cite{Sawicki:2006jj}.
For the DGP model, we can express
$\zeta$ in terms of $\Omega_\b$ and $\Delta$:
\begin{equation}
    \zeta = \Phi + \frac{Ha}{\rho} \delta q =
    -\frac{\dot{H}a^2}{k^2} \Delta
    + \frac{Ha^2}{k^2} \dot\Delta
    -\frac{\epsilon \gamma_1 k^2}{12 Hr_\c a^3} \Omega_\b.
\end{equation}
The derivative of $\zeta$ can then be shown to be
\begin{equation}\label{eq:zeta evolution}
    \zeta' = -\frac{\epsilon}{12} \left(\frac{k}{Ha}\right)^2
    \left( \frac{Hr_\c}{a} \right) \left[ \frac{ \gamma_1 \Omega_\b' + (\gamma_1
    - 9\gamma_3) \Omega_\b }{r_\c^2} \right],
\end{equation}
where a prime indicates differentiation with respect to $\ln a$. One can also confirm that the following identity holds in DGP:
\begin{equation}
\Phi'' - \Psi'  -{H'' \over H'}\Phi' - \left( {H' \over H} -{H''
\over H'} \right) \Psi = -\frac{\epsilon\gamma_1}{12}
\left(\frac{k}{Ha}\right)^2 \left( \frac{Hr_\c}{a} \right)\left[
\frac{\Omega_\b'' - \Omega_\b' + 2H'H^{-2}(\gamma_1
H'+2H)\Omega_\b}{r_\c^2} \right]. \label{eq:zetaconservation}
\end{equation}
Hence, we recover that
\begin{equation}
\zeta' \approx  0 \approx \Phi'' - \Psi'  -{H'' \over H'}\Phi' -
\left( {H' \over H} -{H'' \over H'} \right) \Psi,
\label{eq:zetascaling}
\end{equation}
in the $k \rightarrow 0$ limit \cite{Sawicki:2006jj}.  Given $g$, $\Phi=\Psi (g+1)/(g-1)$ and
this ODE can be readily integrated to find $\Phi$ or $\Psi$ given $g$ and $H$.

Conversely at subhorizon scales $k/Ha \gg 1$, one can employ a ``quasistatic ansatz'' which assumes that gradients with respect to $\mathbf{x}$ and $y$ are of the same order and dominate over time derivatives (see\ \S\ref{sec:QS} for more details on this regime).  This implies that the influence of the bulk through
the brane gradient $R$ is negligible, as it involves only one spatial derivative, and 
\cite{Koyama:2005kd}
\begin{equation}
k^4 \Omega_\b \approx 2 H r_\c {\gamma_4 \over \gamma_3} \kappa_4^2 \rho a^5 \Delta.
\end{equation}
Eq.~(\ref{eq:Phi and Psi}) can be expressed as the Poisson equation \cite{Lue:2004rj}
\begin{eqnarray}
{\Phi-\Psi \over 2} &\approx & {\kappa_4^2 \rho a^2 \over 2k^2} \Delta, 
\label{eq:quasistatic}
\end{eqnarray}
for the quantity $(\Phi-\Psi)/2$ which enters
into observables such as gravitational redshift and lensing.
The quantity $\Psi$, which enters into the motion of non-relativistic matter is
then specified by the relation
\begin{equation}
g \approx g_\text{QS} = -{1 \over 3[1 -2 H r_\c\epsilon(1+ \dot H/3H^2)]}.
\end{equation}

In both limits the dynamics of the perturbations are completely specified
once  $g(a,k)$ is known.    Since $g(a,k)$ is determined in the quasistatic limit by the
structure of the equations themselves, we concentrate on understanding  the superhorizon regime in \S \ref{sec:methodology} and \S \ref{sec:analytic}.   We then seek a suitable interpolation between the two
regimes in \S \ref{sec:comparison}.
This can then be used in the PPF formalism with publicly available codes
 \cite{Fang:2008sn}
to generate the metric perturbations $\Phi$ and $\Psi$  in an efficient manner.

\section{Methodology}
\label{sec:methodology}

In this section we review two techniques, characteristic integration and the scaling ansatz, that have been used in the literature to solve numerically for DGP perturbations on large scales where bulk effects are important.  In the following sections, we will use the scaling ansatz to develop 
the analytic approximations  and the characteristic integration algorithm to test them.

\subsection{Characteristic Integration}

In order to model the behavior of perturbations in the DGP model,
the characteristic integration (CI) algorithm \cite{Seahra:2006tm,Cardoso:2007xc}
constructs a direct finite difference solution to the bulk
(\ref{eq:wave equations}) and brane (\ref{eq:scalar brane equation})
equations of motion subject to the boundary condition
(\ref{eq:scalar boundary condition}).  The code makes use of the
null coordinates $(u,v)$, which implies that the brane is a
\emph{moving} boundary.  

The natural computational domains of the CI
algorithm for both branches are indicated in Figure
\ref{fig:bulkdiagram}.  Initial data for the bulk field $\Omega$
must be supplied on the past null boundary of the domain.  In
addition, one must also specify the value of $\Delta$,
$\dot{\Delta}$ and $\dot{\Omega}_\b$ at the intersection of the
initial data surface and the brane.  Because of the fact that this algorithm is based on null curves in the bulk, the initial condition (\ref{eq:ICs}) is very easy to approximate, we merely need to set $\Omega = 0$ on the initial null surface in the computational domain.  As the initial scale factor of the simulation $a_\text{init}$ is pushed further into the past, the output of the CI code will approach the desired solution with $\Omega(\mathcal{H}_-)=0$, though in practice one finds that simulation results are stable to changes in $a_\text{init}$ provided that it is below some threshold value (typically $10^{-3}$ for the simulations presented in the paper).

\subsection{Scaling Ansatz}
\label{sec:scaling}

During epochs when the
expansion rate on the brane is dominated by a single component,
we expect the master variable on the brane to reach a scaling solution $\Omega_{\rm b}
\propto a^{p}$ for modes that are outside the horizon $(k/Ha) \ll 1$.   The boundary equation~(\ref{eq:scalar boundary condition}) then 
implies that the gradient term of Eq.~(\ref{eq:gradient}) obeys
\begin{eqnarray}
R
&=&-{1 \over 2}\gamma_{1}\left(p^{2}+ {\dot H \over H^{2}} p \right)
+ {9\over 4} \gamma_{3} p - {3 \epsilon\over 4}\gamma_{4} + {3 \over 2} {H r_{c}} { \kappa_{4}\rho
\over H^{2}} \gamma_{4} a^{3}{\Delta\over k^{2}\Omega_{\rm b}} ,
\label{eq:Reqn}
\end{eqnarray}
which gives the relationship between $\Omega_{\rm b}$ and $\Delta$ and
closes the dynamics on the brane.

To determine the gradient parameter $R$ we make a similar scaling ansatz for the
dependence of $\Omega$ on the GN coordinates in the bulk \cite{Sawicki:2006jj}
\begin{equation}\label{eq:scaling ansatz}
    \Omega(t,y) = a^p(t) G(\epsilon H(t)y).
\end{equation}
This
ansatz is motivated by the distance in the bulk connected by a null
worldline $dy = n dt$, i.e. a ``horizon'' in the bulk. For the self-accelerating branch
\begin{equation}
y_{\rm hor} = a H \int_{0}^{a} {d\tilde a \over \tilde a^{2}H^{2}}, \quad (\epsilon=+1),
\end{equation}
and Eq.~(\ref{eq:coordinate transformation}) shows that
$y=y_\text{hor}$ corresponds to the null line $u=0$.
With a
power-law behavior for $H^2 \propto a^{-3(1+w)/2}$, $H y_{\rm hor}
=1/(2+3w)$ if $w > -2/3$. 

For the normal branch, independently of
the evolution of $H$
\begin{equation}
y_{\rm hor} = {1\over H}, \quad (\epsilon=-1),
\end{equation}
which corrects Eq.~(21) in Ref.~\cite{Song:2007wd}.
Eq.~(\ref{eq:coordinate transformation}) shows that $y=y_\text{hor}$
corresponds to the null line $v=0$ on the normal branch and hence
coincides with the coordinate singularity (see discussion below
Eq.~(\ref{eq:Nullbackgroundmetric})).

The final condition introduced in Ref.~\cite{Sawicki:2006jj} is that
the initial data $G(\epsilon H y_{\rm hor})=0$ based on the
assumption that bulk perturbations are generated causally from brane
perturbations.  This is entirely equivalent to Eq.~(\ref{eq:ICs}) and compatible with the CI initial data as discussed above.  Recall that $G(\epsilon H y_{\rm hor})=0$ is also required to keep the Riemann curvature finite in the normal branch.

The scaling technique can be extended to apply
between different scaling regimes by iteratively solving for a time dependent $p(a)$ with
the so-called dynamical scaling (DS) method \cite{Sawicki:2006jj}.  However the CI method
supersedes the DS method in accuracy and so we will not consider the latter further here.
We instead use scaling arguments to develop analytic solutions in the next section.

\section{Analytic Solutions}
\label{sec:analytic}

In this section we employ analytic techniques to obtain solutions for the
metric perturbations modes that are either far inside or outside the horizon on the brane.  
We begin in \S\ref{sec:QS} by deriving the behavior of perturbations in the quasistatic regime $k/Ha \gg 1$, then we examine the opposite large scale limit $k/Ha \ll 1$ for cases where the expansion rate on the brane evolves as a simple power of the
scale factor  in \S \ref{sec:PL}, and finally in \S\ref{sec:dS}-\ref{sec:Green's} we turn to the behavior of superhorizon modes during de Sitter expansion.

\subsection{Quasistatic Modes}\label{sec:QS}

In this subsection, we examine the behavior of modes when $k/Ha \gg 1$ in the context of the quasistatic approximation and derive the first order correction to $g$ in $Ha/k$.  (First order corrections to the quasistatic limit of DGP have also been considered by Amin et al. \cite{Amin:2007wi}.)

If we make the assumption in the master wave equation (\ref{eq:GN wave eqn})
that time derivatives of $\Omega$ can be neglected, we find that  \cite{Koyama:2005kd} 
\begin{equation}
	\Omega \approx c_1 (1+\epsilon H y)^{k/Ha} + c_2 (1+\epsilon H y)^{-k/Ha}
\end{equation}
in the limit of $k/Ha \gg 1$.  To ensure regularity of the master variable on the Cauchy horizon we need to set $c_1 = 0$ for the self-accelerating branch and $c_2 = 0$ for the normal branch.  Therefore, the brane gradient becomes
\begin{equation}
	R = -\frac{\epsilon k}{Ha}\left[ 1 + \mathcal{O}\left( \frac{Ha}{k} \right) \right].
\end{equation}
This analysis justifies the quasistatic assumption in \S \ref{sec:randg} that the impact of
$R$ on the brane is negligible compared with $(k/Ha)^2$ terms in the brane boundary
equation (\ref{eq:scalar boundary condition}).

Putting this into the expressions for the metric potentials (\ref{eq:Phi and Psi}) and expanding in inverse powers of $k/Ha$ we obtain:
\begin{subequations}
\begin{align}
\Phi + \Psi & = \frac{\kappa_4^2 \rho a^2 \Delta}{k^2} \left[ \frac{\epsilon \gamma_1^2}{9Hr_\c \gamma_3} + \frac{\gamma_4(4Hr_\c \gamma_4 - 3\gamma_3)}{3\gamma_3^2} \frac{Ha}{k}  + \mathcal{O} \left( \frac{H^2 a^2}{k^2} \right) \right], \\
\Phi - \Psi & = \frac{\kappa_4^2 \rho a^2 \Delta}{k^2} \left[ 1 - \frac{\gamma_4}{3\gamma_3} \frac{Ha}{k}   + \mathcal{O} \left( \frac{H^2 a^2}{k^2} \right) \right],
\end{align}
\end{subequations}
which in turn yields the first order correction to $g$ from the brane gradient
\begin{equation}
g = g_\text{QS} + \frac{\epsilon \gamma_4 [ \gamma_1^2 + 3\epsilon Hr_\c(4Hr_\c \gamma_4 - 3\gamma_3) ]}{9 Hr_\c \gamma_3^2}  \frac{Ha}{k} + \mathcal{O}\left( \frac{H^2 a^2}{k^2} \right).
\label{eq:qscorrection}
\end{equation}
In Fig.\ \ref{fig:QS first order}, we show a comparison between simulation results and this formula when only the leading order term is retained (i.e.\ the standard quasistatic approximation) and when the next to leading order term is retained (i.e.\ a ``first-order'' quasistatic approximation).  

The first order result gives a more accurate approximation to the simulations results in the $k/Ha \gg 1$ regime. It extends either the accuracy or the regime of validity in
$k/Ha$ of the quasistatic approximation by about an order of magnitude.  
In the example in Fig.\ \ref{fig:QS first order}, it corrects the quasistatic
approximation at the $10^{-3}$ level for $a \le 1$.
 
 For $a \gg 1$, even modes that are far within the current horizon eventually exit the horizon.
Note that the first order correction is unbounded 
as $k/Ha \rightarrow 0$.  The numerical solutions on the self accelerating branch are also
unbounded but the first order correction does not capture this behavior accurately.  
On the normal branch, the numerical solutions remain finite and again the behavior
is not captured by the approximation.  This problem
applies equally to near horizon modes at $a \sim 1$.
We therefore now turn to study the large scale behavior of $g$.

\begin{figure}
\begin{minipage}[t]{0.5\textwidth}
\includegraphics[width=\textwidth]{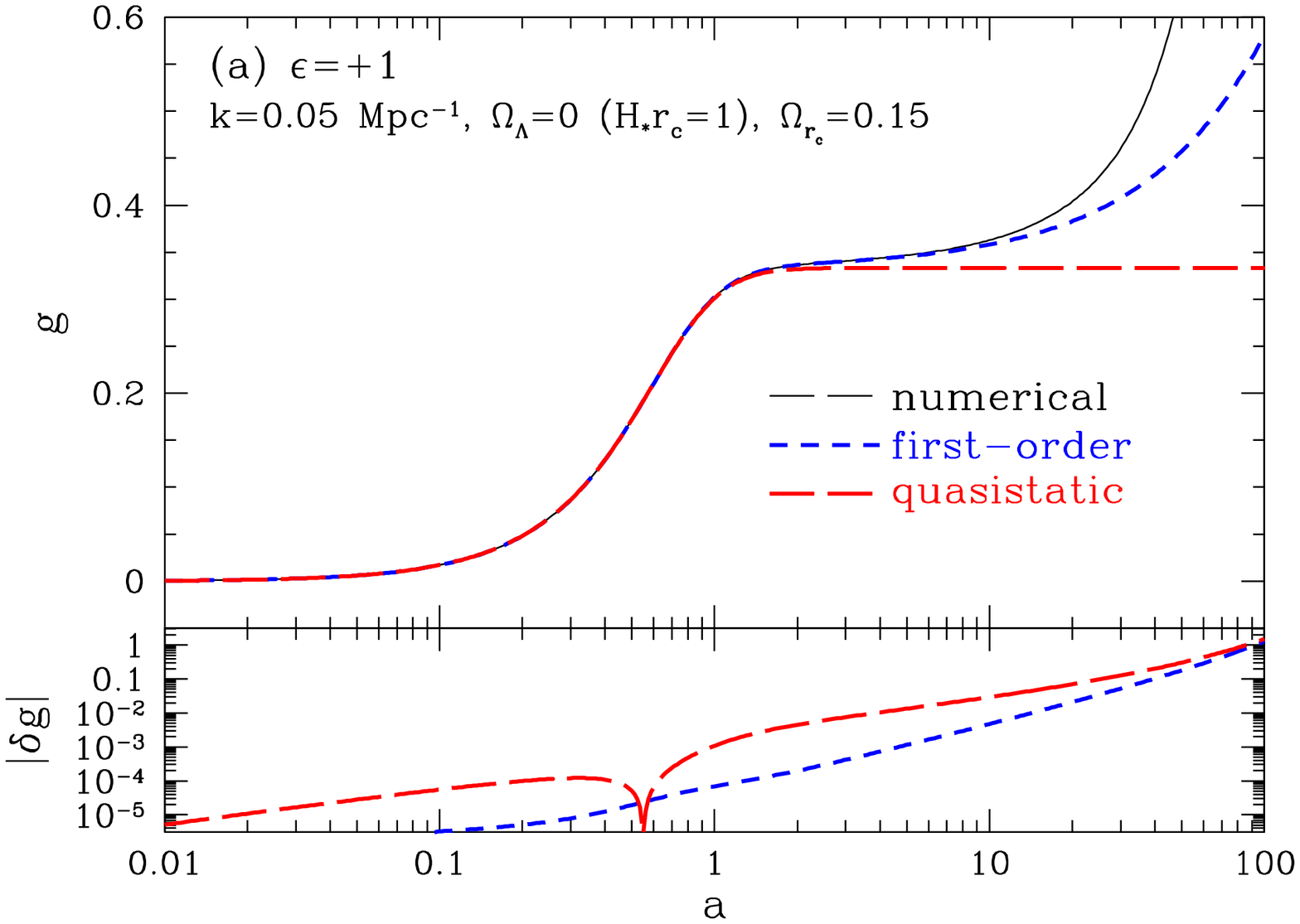}
\end{minipage}%
\begin{minipage}[t]{0.5\textwidth}
\includegraphics[width=\textwidth]{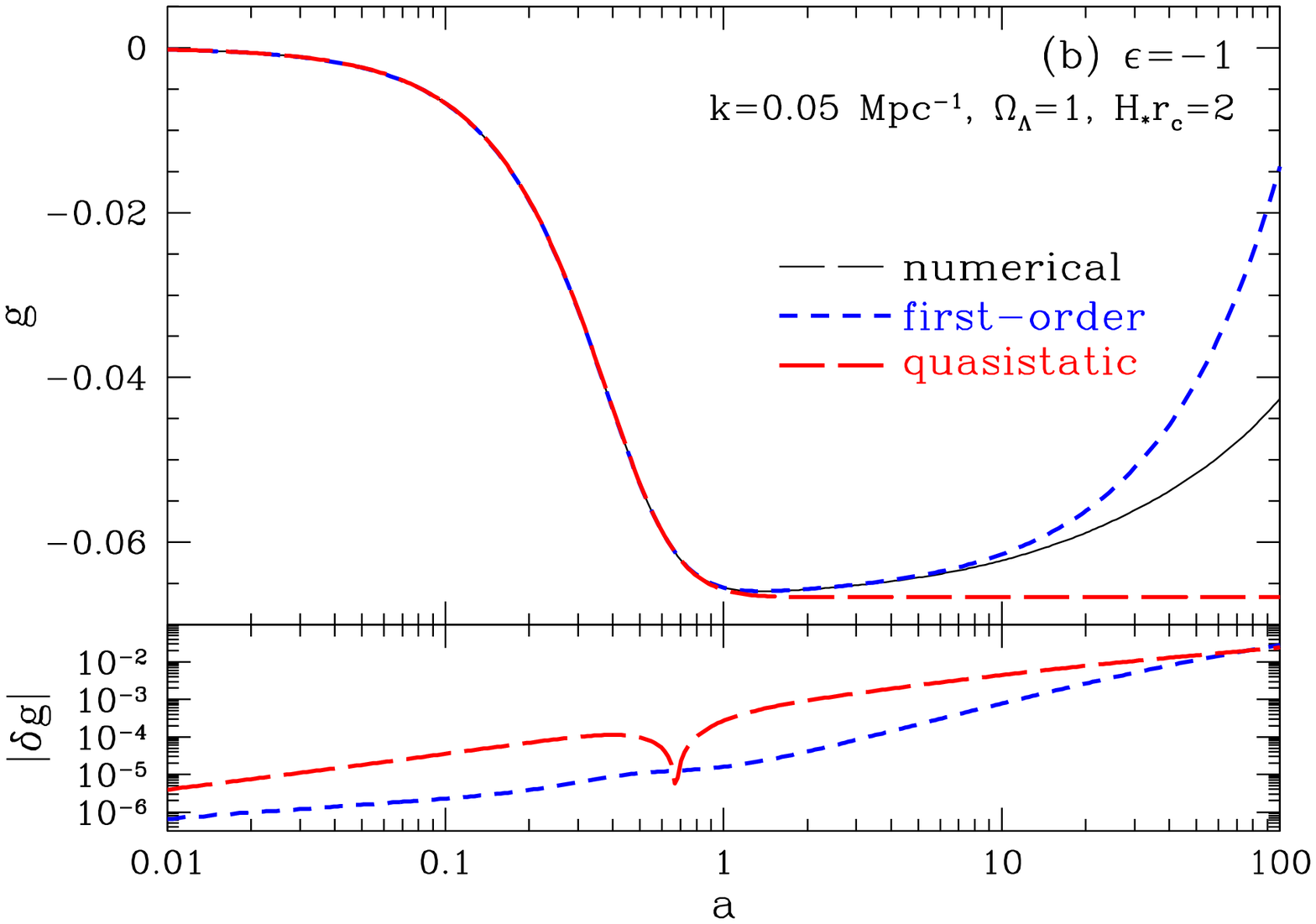}
\end{minipage}
\caption{Comparison of simulation results (black solid) to the quasistatic approximation at zeroth order (red long-dashed) and  with first order $Ha/k$ corrections (blue short-dashed).  In both the self-accelerating and normal branches, the residuals $\delta g$ between the approximation and simulation results are roughly an order of magnitude smaller when first order corrections are included
for $a\le 1$.  
The first order correction becomes invalid as $k/Ha \rightarrow 0$
as the mode exits the horizon during the de Sitter epoch $a \gg 1$.  Also note that we recover the general relativity result
$g = 0$ in the early time limit $Hr_{\c} \rightarrow \infty$ or, equivalently, $a \rightarrow 0$.}\label{fig:QS first order}
\end{figure}

\subsection{Power-law Expansion}
\label{sec:PL}

In the superhorizon regime $k/Ha \ll 1$, there is no universally-valid closed-form solution for $g$.
In this subsection, we thus restrict ourselves to the situation where the
expansion of the brane is well described by a simple power law in the scale factor $a$.  We can then take
\begin{equation}
    H^2 \propto a^{-3(1+w)}, \quad \dot{H} \approx -\frac{3}{2}(1+w)H^2.
\end{equation}
Here, the constant $w$ is an effective equation of state
parameter not to be confused with the equation of state of the matter on the brane. The scaling ansatz (\ref{eq:scaling
ansatz}) reduces the master wave equation (\ref{eq:wave
equations}) down to an ODE for the scaling function $G$ of the form [see \cite{Sawicki:2006jj} Eq.~(43)]
\begin{equation}\label{eq:general G eqn}
    \frac{d^2 G}{dx^2} + \mathcal{P}(x) \frac{dG}{dx} +
    \mathcal{Q}(x) G = 0,
\end{equation}
where $x = \epsilon H y$. The past Cauchy horizon in
the bulk $\mathcal{H}_-$ corresponds to
\begin{equation}
    x_\text{hor} =\epsilon Hy_\text{hor} =
    \begin{cases}
       1/(2+3w), & \epsilon = +1, \, w>-2/3\\
       \text{undefined} , &\epsilon = +1,\,  w\le -2/3\\
       -1, & \epsilon = -1,
    \end{cases}
\end{equation}
while the brane is always at $x = 0$.  The fact that the horizon position is undefined in the self-accelerating case when $w < -2/3$ is a consequence of the behavior of the Gaussian normal coordinates in the bulk.  This is illustrated in Fig.\ \ref{fig:conformal}, where we see that if the effective equation of state $w < -2/3$ at early times, $t =$ constant hypersurfaces fail to intersect $\mathcal{H}_-$ when $\epsilon = +1$.  (This also happens in the normal branch when $w = -1$, but this pathology manifests itself differently below.)  We now seek solutions to
Eq.~(\ref{eq:general G eqn}) in the superhorizon $k/Ha \rightarrow 0$ limit for $x
\in (0,x_\textrm{hor})$ on the self-accelerating branch (when $x_\text{hor}$ is finite) and for $x
\in (x_\textrm{hor},0)=(-1,0)$ on the normal branch.
\begin{figure}
\includegraphics[width=0.95\textwidth]{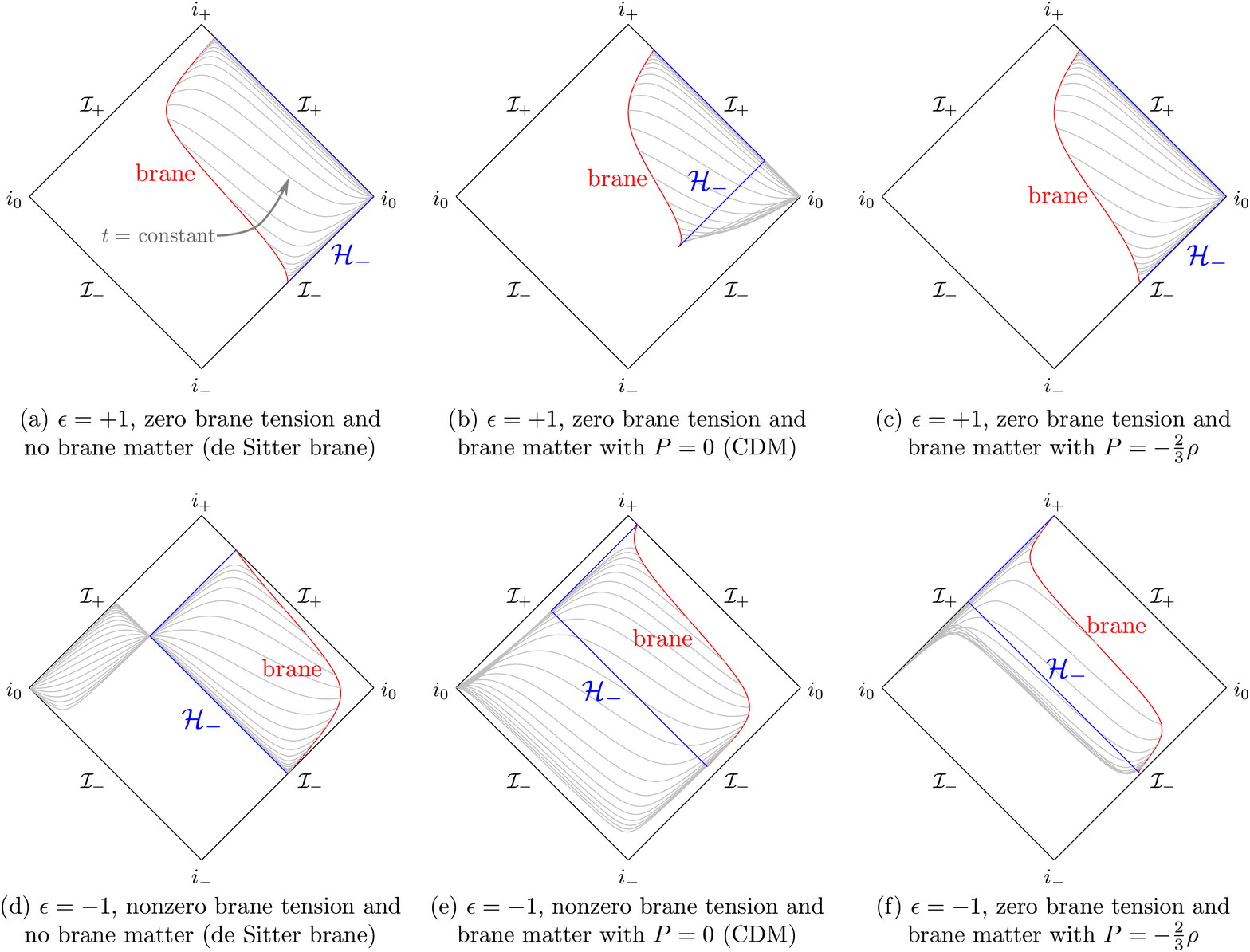}
\caption{Conformal diagrams showing the behavior of the $t =$ constant hypersurfaces 
  in the space spanned by the $(u,v)$ coordinates ($\rho$ and $P$ are the density and pressure of brane matter, respectively).  The scaling ansatz is expected to be informative when these surfaces intersect the brane's past Cauchy horizon $\mathcal{H}_-$.  This is indeed the case in the self-accelerating and normal branches when the brane transitions from early matter domination to a late time de Sitter phase, as in panels (b) and (e) respectively.  However, when the brane undergoes purely de Sitter like expansion, as in (a) and (d), the $t =$ constant curves fail to pierce $\mathcal{H}_-$.  In these scenarios, it is impossible to impose the initial condition $\Omega(\mathcal{H}_-)=0$ in the context of the scaling ansatz.  Finally, there are some choices of brane matter and tension where the behavior of the $t =$ constant surfaces are qualitatively different on the two
branches.  An example of this is in panels (c) and (f), where the brane matter has $P = -\frac{2}{3}\rho$.  In the normal branch $t =$ constant surfaces cover $\mathcal{H}_-$ and we expect the scaling calculation to be applicable, while the opposite is true for the self-accelerating case.}\label{fig:conformal}   
 \end{figure}

A straightforward analysis of the poles of $\mathcal{P}$ and
$\mathcal{Q}$ reveals that $x = 1/(2+3w)$ and $x = 2/(3w+1)$ are
regular singular points of the ODE (\ref{eq:general G eqn}).  The
first of these is coincident with the bulk horizon when $\epsilon =
+1$ and $w>-2/3$.  The second singularity occurs outside the range $x \in
[-1,1/(2+3w)]$ for all $w$, and hence is not relevant to our
problem.

 In addition to the regular singular points, there is an
irregular singular point at $x = -1$; i.e., coincident with the bulk
horizon when $\epsilon = -1$.  The nature of this singularity is
apparent if we expand $\mathcal{Q}$ in a Laurent series about $x =
-1$:
\begin{equation}\label{eq:Laurent}
    \mathcal{Q}(x) = -\frac{3}{4} \frac{k^2}{H^2 a^2}
    \frac{1+w}{(1+x)^3} + \cdots
\end{equation}
We see that this third order pole disappears if either $w = -1$ or
$k/Ha = 0$.  

We use the technique of matched asymptotic expansion to obtain the solution in
the $k/Ha \ll 1$ limit.   Away from the pole at $x=-1$, we can set $k=0$ in
Eq.~(\ref{eq:general G eqn})  to obtain the
outer (or near-brane) solution
valid for $w \ne -2/3$
\begin{equation}
G(x) = c_1 (1+x)^p + c_2 (1+x)^{3/2} [1 - (2+3w)x]^{(2p-3)/(4+6w)}.
\label{eq:Ggeneralw}
\end{equation}
Here, $c_1$ and $c_2$ are arbitrary constants.

The outer solution is sufficient for the consideration of the self accelerating branch
where $x \in (0, x_{\rm hor})$.  When $w>-2/3$,
 the condition that
$\Omega(\mathcal{H}_-) = 0$ is equivalent to setting
$G(1/(2+3w))=0$, which implies that $c_1 = 0$ when $p > 3/2$.
Under these conditions, the brane gradient is
\begin{equation}
    R = \left( \frac{1}{G}\frac{dG}{dx} \right)_{x=0} = 3-p.
\end{equation}
For $p<3/2$, strictly speaking no solution exists for constant $w$.  However in the real universe, one would expect deviations
from other components that break the $w=$const. assumption to make the $c_2$ solution finite at the horizon and $c_1$ dominant at 
the brane.  Hence we can assign $R=p$ to
the $p<3/2$ modes.

For $w<-2/3$ as in the case of the de Sitter regime considered below,
$x_\text{hor}$ in principle diverges.  In practice $x_\text{hor}$ always remains finite
at finite time given the preceding epochs of radiation and matter domination but increases
without bound.

Turning our attention to the normal branch, we require an
inner (or near-horizon) solution close to the irregular singular point at $x=-1$.
The
Laurent expansion (\ref{eq:Laurent}) implies that we cannot set $k =
0$ \emph{a priori}; i.e., the $k/Ha$ contributions to the ODE are
not subdominant near $x = -1$.  To overcome this, we transform to
the variable
\begin{equation}
    \xi = \frac{Ha}{k} \sqrt{\frac{1+x}{3(1+w)}} . 
    \end{equation}
In terms of this new coordinate, the horizon is at $\xi = 0$ and the
brane is at $\xi\gg 1$.    The inner solution is specified by setting $k=0$
in the transformed equation for $\xi$ and takes the form
\begin{equation}\label{eq:near horizon G}
    G(x(\xi)) = \xi^{p+3/2} \left[ c_3 I_{3/2-p} \left( \frac{1}{\xi}
    \right) + c_4 K_{p-3/2} \left( \frac{1}{\xi}
    \right) \right],
\end{equation}
where $I$ and $K$ are modified Bessel functions of the first and
second kind, respectively.

On the normal branch, the boundary condition that $\Omega(\mathcal{H}_-) = 0$ implies
$G = 0$ at $\xi = 0$.  Taking note of the asymptotic expansions of
the Bessel functions for $\xi \ll 1$,
\begin{equation}
    I_{3/2-p} \left( \frac{1}{\xi}
    \right) \sim \exp \left( \frac{1}{\xi}
    \right), \quad K_{p-3/2} \left( \frac{1}{\xi}
    \right) \sim \exp \left( - \frac{1}{\xi}
    \right),
\end{equation}
we see that we must set $c_3 = 0$ in the near-horizon $G$ solution
(\ref{eq:near horizon G}).

Finally we match the $c_4$ inner  solution (\ref{eq:near horizon G}) to the outer solutions (\ref{eq:Ggeneralw}) at $\xi \gg 1$ but $(1+x) 
\ll 1$ to
obtain the global approximate solution
\begin{equation}
G(x(\xi)) \propto
\begin{cases}
 \xi^{p+3/2} K_{p-3/2}(1/\xi), & p \ge 3/2, \\
 \xi^{p+3/2} K_{p-3/2}(1/\xi) [1-(2+3w)x]^{(2p-3)/(4+6w)}, & p < 3/2.
  \end{cases}
\label{eq:global}
\end{equation}
Substituting this global solution into the exact equation (\ref{eq:general G eqn}) for $G$, we see that the residuals are ${\cal O}(k/Ha)$ and converge
uniformly in the interval $0 < 1+x< 1$; moreover numerical integration of
Eq.~(\ref{eq:general G eqn}) confirms these solutions.
Given  $\xi^2 \propto (1+x)$, this solution implies that $p\ge 3/2$ matches onto
the $c_1$ mode $p<3/2$ the $c_2$ mode of Eq.~(\ref{eq:Ggeneralw}) for the normal
branch.  Before moving on, we note that these expressions for $G$ in the normal branch only hold when the irregular singular point at $x = -1$ is present; i.e.~when $w > -1$ and $k$ is finite.  In other words, the scaling approximation will not be informative in the purely de Sitter case $w = -1$, as could have been surmised from panel (d) in Fig.\ \ref{fig:conformal}.

To summarize, we have shown that under the scaling
ansatz $\Omega(t,y) = a^p G(\epsilon H y)$ combined with the initial data
$G(\epsilon H  y_{\rm hor})=0$ the brane
gradient satisfies
\begin{equation}
    R = \begin{cases}
        3 -p, & \epsilon = +1, \, w>-2/3, \, p\ge3/2, \\
        p, & \epsilon = +1, \, w>-2/3, \, p<3/2, \\
        p, & \epsilon = -1,  \,  w\ne -1, \, p\ge3/2,\\
        3-p, & \epsilon= -1, w\ne -1, \, p<3/2
    \end{cases}
    \label{eq:Rgenw}
\end{equation}
in the $k/Ha \rightarrow 0$ limit.

The preferred values for $p$ are determined by the brane boundary equation through
the density source.  For example in the matter dominated phase when $w=0$, the matter density fluctuation $\Delta \propto a$ acts
as an external source to $\Omega_\b$ in the boundary equation
(\ref{eq:scalar boundary condition}) so that for $k/Ha \ll 1$, the particular mode is
$p=4$.
For the self accelerating branch the matter dominated solution is therefore $R=-1$ and
\begin{equation}
g_{\rm SH}  = { 9 \over 8 H r_\c -1}, \quad \epsilon =+1 ,
\end{equation}
whereas on the normal branch $R=4$ and
\begin{equation}
 g_{\rm SH}  = - { 1 \over 2 H r_\c +1}, \quad \epsilon =-1 .
\end{equation}
These relations close the perturbation equations on the brane.

\subsection{De Sitter Scaling}
\label{sec:dS}
%\subsubsection{The de Sitter approximation}

The late time de Sitter phase carries an effective equation of state of $w=-1$
and is a special case for the scaling solutions both on the self-accelerating and normal branch.  The horizon
on the self-accelerating branch goes to infinity while the horizon on the normal branch is no longer an irregular singular point of the 
master equation~(\ref{eq:GN wave eqn}).

Given the matter in the universe, a pure de Sitter expansion is never fully reached and
it is important to consider the manner in which the de Sitter phase is approached.
The Hubble parameter approaches a constant
\begin{equation}
{H_\star \over H_0} = \sqrt{\Omega_\Lambda + \Omega_{r_\c}}+ \epsilon \sqrt{\Omega_{r_\c}}
\label{eq:Hstar}
\end{equation}
for a scale factor $a \gg a_\star$ where $H(a_\star) \equiv \sqrt{2} H_\star$ and
\begin{equation}
\Omega_m a_\star^{-3} = \Omega_\Lambda + (4-2\sqrt{2}) \Omega_{r_\c} (1+ \epsilon
\sqrt{1 + \Omega_\Lambda/\Omega_{r_\c}}).
\label{eq:astar}
\end{equation}
Note that if $\Omega_\Lambda/\Omega_{r_\c} \gg 1$, $a_\star = (\Omega_m/\Omega_\Lambda)^{1/3}$ as usual.
If $\Omega_\Lambda/\Omega_{r_\c} \ll 1$
\begin{equation}
a_\star =
\begin{cases}
\displaystyle \left[ {\Omega_m \over 4 (2-\sqrt{2})\Omega_{r_\c}} \right]^{1/3}, & \epsilon =+1, \\
\displaystyle \left[ {\Omega_m \over (\sqrt{2}-1)\Omega_\Lambda} \right]^{1/3}, & \epsilon =-1. \\
\end{cases}
\end{equation}
In both cases $H r_\c$ also approaches a constant
\begin{equation}
    H_\star r_\c = \frac{1}{2}\left[ \sqrt{1 +
    \frac{\Omega_\Lambda}{\Omega_{r_\c}}}
    + \epsilon \right].
\end{equation}
We shall see that the residual effects from the preceding matter dominated phase
determine the behavior of the perturbations on the brane.
Likewise in Fig.~\ref{fig:conformal}, we see that the pathologies in panels
(a) and (d) disappear with the addition of matter in (b) and (e).

\subsubsection{Bulk solutions}\label{sec:scaling de Sitter}

If we take a pure de Sitter limit where $\dot H=0$,
the master equation (\ref{eq:GN wave eqn}) becomes
\begin{equation}
(1+x)^2{\di^2 \Omega \over \di x^2} -2 (1+x){\di \Omega \over \di x} -\Omega'' + 3\Omega'
- {k^2 \over a^2 H^2}\Omega = 0 ,
\label{eq:masterpuredS}
\end{equation}
where recall $' = d/d\ln a$.

The general solution to the pure de Sitter master equation for $k/Ha \ll 1$ is
\begin{equation}
\Omega = a^p [ c_1 (1+x)^{p} + c_2 (1+x)^{3-p}].
\label{eq:deSkzero}
\end{equation}
This solution in fact corresponds to the $w=-1$ limit of the general constant $w$
scaling solution
(\ref{eq:Ggeneralw})
but in this case separability in $a$ and $x$ is guaranteed, not an ansatz.

One might therefore expect the scaling results to hold:
 $c_1=0$ ($R=3-p$) for the self-accelerating branch and $c_2 = 0$ ($R=p$)
  for the normal branch for the
 fastest growing modes. However, direct application of the condition $G(x_{\rm
hor})=0$ fails to be informative in both cases due to a change in the
nature of the $x_{\rm hor}$ point. 
  For the self-accelerating branch, its
value diverges and for the normal branch it is no longer an irregular
singular point of the master equation.  The can also be seen in Fig.\ \ref{fig:conformal}, where we see that if the brane undergoes a purely de Sitter expansion, the $t =$ constant hypersurfaces do not intersect the past horizon and it is impossible to impose the initial condition $\Omega(\mathcal{H}_-) = 0$ in the scaling approximation.

To restore the information lost in the pure de Sitter limit, we must
consider the preceding matter dominated expansion as in panels (b) and (e) of Fig.\ \ref{fig:conformal}.  For the
self-accelerating branch, the matter dominated phase leads to a
finite but growing $x_{\rm hor} \sim a H_\star^2/\Omega_m H_0^2$.
The $c_1$ solution however does not depend on the evolution of
$H(a)$ and so even given this preceding phase its contribution to
$\Omega$ at the horizon increases as $(1+x_{\rm hor})^p$ for $p>0$.
Hence the $c_1$ contribution must be negligible near $x\sim 0$ in
order to satisfy the initial data $G(x_{\rm hor})=0$.   The form of
the $c_2$ solution does depend on the evolution of $H(a)$ and hence
it is plausible that the preceding matter dominated epoch induces a
correction of the form $(1-x/x_{\rm hor})$ as it does in the matter
dominated limit [see Eq.~(\ref{eq:Ggeneralw})].  This correction has
negligible impact near $x=0$ and so we again obtain
\begin{equation}
R=3-p, \quad \epsilon = +1,
\end{equation}  for the solution on the self-accelerating
branch.  We shall see that this line of reasoning is borne out by the more detailed
Green's function calculation below.

On the normal branch,
restoring a trace amount of matter has more dramatic effects.  Defining
\begin{equation}
h_m \equiv { H^2 - H_{\star}^2 \over H_\star^2} \propto a^{-3},
\end{equation}
we obtain $H'/H \approx -(3/2) h_m$.  In Eq.~(\ref{eq:GNbackgroundmetric}), the zero of
$n$ is shifted beyond the horizon $x<-1$ and the master equation (\ref{eq:GN wave eqn})
regains an irregular singular point at finite $k/Ha$.

We can transform to the variable
\begin{equation}
    \xi = \frac{Ha}{k} \sqrt{\frac{1+x}{3 h_m}}
\end{equation}
and again repeat the asymptotic matching of the interior $k=0$ and exterior $k=0$
solutions.   As with constant $w$ the interior solutions are given by
Eq.~(\ref{eq:near horizon G}).   For $p>3/2$ they again match onto the $c_1$ solution with
$R=p$.  For $p<3/2$, it is important to consider an intermediate solution at finite
$h_m$ but $k=0$.    The interior solution matches onto the $(1+x)^{3/2}$ rather than the
$(1+x)$ piece of this solution and in general can stimulate both the $c_1$ and the
$c_2$ pieces of the exterior solution.
From explicit solutions of these equations, we find that as $h_m \rightarrow 0$ the $c_2$ piece
where $R=3-p$ generally dominates given its larger growth with $(1+x)$.   We shall see however that
this leading order behavior is forbidden by the brane boundary equation.

\subsubsection{Brane boundary}

As in the matter dominated limit, certain values of  the scaling index $p$ are picked out by the
boundary equation (\ref{eq:Reqn}).
Once the matter contribution to the expansion becomes
negligible $\kappa_4^2 \rho /H^2 \ll 1$ and the matter conservation law
(\ref{eq:scalar brane equation}) requires that density perturbations obey
$\Delta=\Delta_0 + \Delta_1 (a/a_\star)^{-2}$ where $\Delta_0$ and $\Delta_1$ are constants, i.e.~to leading order, density 
fluctuations freeze out but with some stimulation
of a decaying mode at the de Sitter transition.

While this behavior of $\Delta$ is the same as in General Relativity with dark energy,
 the particular mode here does not drive the leading order behavior
of the metric perturbations $\Phi$ and $\Psi$.   Note that the boundary equation
(\ref{eq:scalar boundary condition}) and
master equation
(\ref{eq:GN wave eqn})
 requires a constant response in $\Omega_\b$
\begin{equation}
\Omega_\b = {2 \epsilon H r_\c \kappa_4^2 \rho a^3 \over k^2 H^2}\Delta_0,
\label{eq:scalingparticular}
\end{equation}
where the specific coefficient applies to $R=p=0$.    By virtue
of Eq.~(\ref{eq:Phi and Psi}), this mode caries no source to $\Phi$
and $\Psi$ and as we shall see is the relevant particular mode for the normal branch [see Eq.~(\ref{eq:particularscaling})].  Hence their evolution  is typically governed instead by the  homogeneous $\Delta$ source free modes in $\Omega_\b$.
If the source free modes decay faster than
$p=-2$, then the next to leading order scaling of $\Delta_1$
will dominate instead.  On the self accelerating branch, we shall see that the fastest growing modes have $p>0$ and again the leading order particular contribution does not dominate
the metric evolution.

Now let us consider these homogeneous modes.
We obtain from Eq.~(\ref{eq:Reqn})
\begin{equation}
p(p-3) + {1 \over \epsilon H r_\c} + {2 \epsilon H r_\c - 1 \over \epsilon H r_\c} R = 0 .
\label{eq:ptoRdS}
\end{equation}
For the self-accelerating branch, the condition $R=3-p$ ($c_1=0$) gives
\begin{equation}
p = 2, \, 3-{1 \over H r_\c} , \quad \epsilon = +1,
\end{equation}
where note that if $\Omega_\Lambda=0$, $H r_\c \rightarrow 1$ and
both solutions return $p=2$.    This analysis justifies the
selection of the $p=2$ mode in \cite{Sawicki:2006jj}.

For the normal branch, taking $R=p$ ($c_2=0$) gives
\begin{equation}
p = 1, \, -{1 \over H r_\c}, \quad \epsilon = -1
\label{eq:pnormal}
\end{equation}
Here $p<3/2$ and so in principle the  $R=3-p$ mode should dominate leading to a
contradiction. If we instead assume $R=3-p$, Eq.~(\ref{eq:ptoRdS}) requires that either
$p = 2$ or $p = 3+1/Hr_\c $; in either case, $p>3/2$ and we again reach a contradiction.
Therefore,  no solution of this type can satisfy both $G(x_{\rm hor})=0$
and the brane boundary condition.

Since the details of the matter to de Sitter transition
are important for the $R=3-p$ mode whereas they are not for the $R=p$ mode, we
conjecture  that solutions with $p<3/2$ should exist if corrections near the singularity
 prevent the $c_2$ mode in Eq.~(\ref{eq:deSkzero}) from completely dominating.

 If we further assume that there exist modes where $R=p$ is the dominant mode at $p<3/2$
 we obtain the modes in Eq.~(\ref{eq:pnormal}). In principle though, we can obtain a spectrum
 of modes with mixed $R=p$ and $R=3-p$ behavior.
 We shall see that the numerical CI solution implies that $p=1$ is indeed the dominant mode for
 $\Omega$.    However, this mode has no effect on the brane metric parameters $\Phi$ and $\Psi$.
Hence the important  $k/Ha \rightarrow 0$ mode is $p=-1/H r_\c$, but that is a
decaying mode in $\Omega$.  This opens up the possibility that $k$-dependent modes
eventually dominate the evolution in the de Sitter epoch.   To study this behavior and
verify the conjectures involving the $c_2$ mode for both the self-accelerating and normal
branch,
we require a more complete Green's function analysis.

\subsection{De Sitter Green's Function}\label{sec:Green's}

To gain further insight into solutions in the de Sitter epoch,
we can recast the perturbation equations as a canonical
scattering problem via the transformation
\begin{equation}\label{eq:phi defn}
    \Omega(t,y) = (1+\epsilon \Hinf y)^{3/2} a^{3/2}(t)
    \varphi(t,y),
\end{equation}
and define a new bulk variable $z$:
\begin{equation}
    z = \frac{\ln(1+\epsilon \Hinf y)}{\epsilon \Hinf}, \quad
    y = \frac{e^{\epsilon \Hinf z} -1}{\epsilon \Hinf}.
\end{equation}
The bulk manifold is defined by $z \in [0,\infty)$ for both
branches.  The new bulk variable satisfies a simple wave equation
and boundary condition
\begin{subequations}\label{eq:canonical system}
\begin{gather}\label{eq:simple wave equation}
    \left(\frac{\di^2}{\di t^2} - \frac{\di^2}{\di z^2} + \frac{k^2}{a^2} \right) \varphi =
    0, \\ \label{eq:simple boundary condition}
    \left( \frac{\di \varphi}{\di z} \right)_{z=0} = -\frac{\epsilon r_\c}{2\Hinf r_\c
    -\epsilon} \left( \frac{\di^2\varphi}{\di z^2} \right)_{z=0} -
    \frac{\epsilon \Hinf (3\Hinf r_\c - 2\epsilon)}{4(2\Hinf r_\c -
    \epsilon)} \varphi_\b + \frac{2\epsilon\kappa_4^2 r_\c \rho a^{3/2}}
    {k^2(2\Hinf r_\c -\epsilon)} \Delta,
\end{gather}
\end{subequations}
where we have set $\varphi_\b(t) = \varphi(t,0)$ and made use of
Eq.~(\ref{eq:simple wave equation}) to remove time derivatives of
$\varphi$. The equation of motion for $\Delta$ reduces to
\begin{equation}\label{eq:Delta EOM}
    \ddot\Delta + 2\Hinf \dot\Delta -
    \frac{\kappa_4^2 \rho \Hinf r_\c}{2\Hinf r_\c - \epsilon} \Delta =
    \frac{\epsilon k^4}{3(2\Hinf r_\c - \epsilon) a^{7/2}} \varphi_\b.
\end{equation}
Examining equations (\ref{eq:canonical system}), we see that
$\Delta$ can be viewed as a brane-localized source for the bulk
$\varphi$ field.  From linearity, it follows that the general
solution for $\varphi$ should be of the form
\begin{equation}
    \varphi = \varphi^\homo + \varphi^\particular,
\end{equation}
where $\varphi^\homo$ is the general homogeneous solution for
Eq.~(\ref{eq:canonical system}) when $\Delta = 0$, and
$\varphi^\particular$ is a particular solution when $\Delta$ is
nonzero and determined from Eq.~(\ref{eq:Delta EOM}).

Let us now concentrate on the solution of the homogeneous system
where $\varphi^\homo$ and its first time derivative
$\dot\varphi^\homo$ are known at an initial time, say $t = 0$. Then,
the value of $\varphi^\homo$ at some later time $t$ and position $z$
is given by
\begin{multline}\label{eq:IVP soln}
    \varphi^\homo(t,z) = \int_0^\infty dz'\, \left[ \varphi^\homo(t',z')
    \frac{\di}{\di t'}G_-(t,t';z,z') - G_-(t,t';z,z') \frac{\di}{\di t'}
    {\varphi^\homo} (t',z') \right]_{t'=0}
    + \\
    \frac{\epsilon r_\c}{2\Hinf r_\c - \epsilon} \left[ \varphi^\homo(t',z')
    \frac{\di}{\di t'}G_-(t,t';z,z') - G_-(t,t';z,z')\frac{\di}{\di t'}{\varphi^\homo} (t',z')
    \right]_{{t'=0},{z'=0}},
\end{multline}
where $G_-$ is the retarded Green's function.  Though this formal solution of the initial value problem only involves $G_-$, we will also consider the advanced Green's function $G_+$ to facilitate comparison with the scaling solutions of \S\ref{sec:dS}.   Both Green's functions satisfy
\begin{equation}\label{eq:Gsimple wave equation}
    \left(\frac{\di^2}{\di t^2} - \frac{\di^2}{\di z^2} + \frac{k^2}{a^2} \right) G_\pm(t,t';z,z') =
    -\delta(t-t')\delta(z-z'),
\end{equation}
with boundary conditions
\begin{subequations}
\begin{gather}
    0 = \left\{ \left[ \frac{\di}{\di z_<} + \frac{\epsilon r_\c}{2\Hinf r_\c
    -\epsilon} \frac{\di^2}{\di z_<^2} +
    \frac{\epsilon \Hinf (3\Hinf r_\c - 2\epsilon)}{4(2\Hinf r_\c -
    \epsilon)} \right] G_\pm(t,t';z,z') \right\}_{z_<=0}, \label{eq:Green's brane BC} \\
    0 \le \Delta z \pm \Delta t \quad \Rightarrow \quad G_\pm(t,t';z,z') = 0, \label{eq:Green's far BC}
\end{gather}
\end{subequations}
where $z_< = \min(z,z')$, $z_> = \max(z,z')$, $\Delta t = t - t'$,
and $\Delta z = |z - z'|$.  The boundary condition (\ref{eq:Green's far BC}) ensures that the initial condition $\Omega(\mathcal{H}_-) = 0$ will be satisfied by Eq.~(\ref{eq:IVP soln}) for $t >0$.  Note that $\mathcal{H}_-$ corresponds to $t \rightarrow -\infty$ and $z \rightarrow \infty$, yet the boundary condition (\ref{eq:Green's far BC}) is imposed at finite values of the coordinates.  This is in contrast to the scaling approach, where the condition $\Omega = 0$ is enforced on $\mathcal{H}_-$ directly. Therefore, unlike the scaling solution, the Green's function approach is not compromised by the fact that Gaussian normal coordinates do not cover the past horizon for a de Sitter brane. That is, it is not necessary to impose conditions on the Green's function at $\mathcal{H}_-$ to ensure that the solution (\ref{eq:IVP soln}) is consistent with $\Omega(\mathcal{H}_-) = 0$ .

Using separation of variables and other standard techniques, the
advanced and retarded Green's functions can be expressed as
\begin{subequations}
\begin{gather}\label{eq:Green funcn solution}
    G_\pm(t,t';z,z') = \int_{-\infty}^{+\infty} d\nu \,
    \mathcal{G}_\pm(t,t';z,z'|\nu), \\
    \mathcal{G}_\pm(t,t';z,z'|\nu) = \frac{T_\nu(t) T_\nu^*(t')
    [f_\mp(\nu) e^{\pm i\nu \Hinf |z-z'|}+f_\pm(\nu)
    e^{\pm i\nu\Hinf(z+z')}]}{2i\nu \Hinf f_\mp(\nu)},
\end{gather}
\end{subequations}
The two terms represent propagation directly from $z'$ to $z$ or through reflection off
of the boundary.  The functions $T_\nu$ satisfy
\begin{equation}
    \left(\frac{\di^2}{\di t^2} + \frac{k^2}{a^2} + \Hinf^2 \nu^2 \right)
    T_\nu(t) =0, \quad \int_{-\infty}^{+\infty} d\nu \, T_\nu(t)
    T_\nu^*(t') = \delta(t-t'),
\end{equation}
and $f_\pm$ follow from the boundary condition (\ref{eq:Green's
brane BC}):
\begin{equation}
    f_\pm(\nu) = (2\epsilon\nu \pm i)(3i\Hinf r_\c - 2i\epsilon \pm 2\epsilon\Hinf r_\c
    \nu).
    \label{eq:fpm}
\end{equation}
Explicitly, the temporal mode functions are given by
\begin{equation}
    T_\nu (t) = \left( \frac{k}{2\Hinf} \right)^{i\nu}
    \left( \frac{\Hinf\nu}{\sinh \pi \nu} \right)^{1/2} J_{-i\nu}\left( \frac{k}{a\Hinf}
    \right).
\end{equation}
Here, $J_{-i\nu}$ is the Bessel function of the first kind with
order $-i \nu$.

\begin{table*}
\newcolumntype{Q}{>{$}m{32mm}<{$}}
\begin{tabular}{QQQQQ}
\hline \text{Green's function} & \text{resonant mode}
& \text{frequency } \nu & \text{temporal scaling } p & \text{brane gradient } R \\
\hline \hline%
 \text{retarded } G_- & A_1 & -i\epsilon/2 & (3+\epsilon)/2 &
1 \\
& A_2 & -i(3\epsilon/2-\omega) & 3(1+\epsilon)/2 -
\omega & \epsilon\omega \\
& B_m \,\, (m=1,2\ldots) & im & 3/2 - m &
\gamma_1(9-4\epsilon \omega-4 m^2)/8
\\ \hline

\text{advanced } G_+ & C_1 & i \epsilon/2
& (3-\epsilon)/2 & 1 \\
& C_2 & i ( 3\epsilon/2-\omega) & 3(1-\epsilon)/2 +
\omega & \epsilon \omega \\
& D_m \,\, (m = 1,2\ldots) & im & 3/2 - m & \gamma_1(9-4\epsilon\omega-4 m^2)/8 \\

\hline
\end{tabular}
\caption{Resonances of the retarded and advanced Green's functions as determined by the poles of $\mathcal{G}_-$ and $\mathcal{G}_+$, respectively.
($\omega = 1/\Hinf r_\c$).  Note that the $A_{1,2}$ and $C_{1,2}$
excitations give rise to late time resonant modes of the form
$\Omega \propto a^p (1+\epsilon \Hinf y)^{R}$, while the
$B_m$ and $D_m$ modes have more complicated $y$ dependence.} \label
{tab:resonances}
\end{table*}

Now, in order to evaluate the righthand side of Eq.~(\ref{eq:Green
funcn solution}), we close the contour of integration in the
upper-half of the complex $\nu$ plane when $\mp \Delta t > \Delta
z$, and in the lower-half plane otherwise. This gives
\begin{equation}\label{eq:asymptotic Green's}
    G_\pm(t,t';z,z') = \Theta(\mp \Delta t - \Delta z) \left[ 2\pi i \sum \text{Res} \, \mathcal{G}_\pm(t,t';z,z'|\nu) -
    \int_{\Gamma_+} d\nu \, \mathcal{G}_\pm(t,t';z,z'|\nu) \right].
\end{equation}
In this expression, the sum is over the residues of the poles of
$\mathcal{G}_\pm(t,t';z,z'|\nu)$ in the upper-half of the complex
$\nu$ plane, and the contour $\Gamma_+$ represents the infinitely
large semi-circle used to close the integration path.  The poles of
$\mathcal{G}_\pm(t,t';z,z'|\nu)$ represent resonant excitations that
will dominate the late time behavior of our system, and are listed
in Table \ref{tab:resonances}. The $A$ and $C$ resonances listed in the table arise from zeros in $f_\pm(\nu)$ in Eq.~(\ref{eq:fpm}), while the $B$ and $D$ modes are associated with divergences in the $T_\nu(t)$ functions.  In what follows, we will neglect the
$\Gamma_+$ integration and represent the Green's function as a sum
over resonances.

It is instructive to relate the $A$ and $C$ resonances to the preferred modes in the scaling approach. The scaling
solution of Eq.~(\ref{eq:deSkzero}) corresponds to
\begin{equation}
\varphi(t,y) = c_1 e^{(p-3/2)\Hinf(t+\epsilon z)} + c_2 e^{(p-3/2)\Hinf(t-\epsilon z)}.
\label{eq:scalingvarphi}
\end{equation}
Each term represents a wave traveling towards or away from the brane depending on the choice of branch, and the boundary equation satisfied by $\varphi$ can be interpreted as a reflection condition on these waves. Comparing the resonances from the poles of $\mathcal{G}_\pm$ to Eq.~(\ref{eq:scalingvarphi}), we see that the $A_{1,2}$ resonances of the \emph{retarded} Green's function correspond to scaling solutions with $c_1 = 0$ in the self-accelerating branch and $c_2 = 0$ in the normal branch.  Conversely, the $C_{1,2}$ \emph{advanced} resonances have $c_2 = 0$ and $c_1 = 0$ for the self-accelerating and normal branches, respectively.   The causality condition (\ref{eq:ICs}) picks out the retarded solution in each branch, which in turn implies that $c_1=0$ and $R =  3 - p$ for the self-accelerating branch, while $c_2=0$ and  $R = p$ for the normal branch.  This analysis verifies our conjectures of the previous section based on extending the de Sitter scaling solutions to include trace amounts of matter to specify the initial data.

In order to facilitate the comparison of simulation results with the
predictions of the Green's function analysis, we hereafter restrict our
attention to the retarded Green's function and assume that the field
point $(t,z)$ is on the brane and deep in the de Sitter era; i.e.\
$k/\Hinf a(t) \ll 1$. The asymptotic form of $G_-$ in this limit is
\begin{equation}
    G_-(t,t';0,z') \approx -\frac{8i\epsilon \Hinf r_\c}{\gamma_1\pi} \int_{-\infty}^{+\infty} d\nu \,
    \left[ \frac{k}{2\Hinf a(t)} \right]^{-i\nu} \frac{\nu
    \Gamma(i\nu)}{f_+(\nu)} e^{- i\nu \Hinf z'} J_{i\nu} \left[
    \frac{k}{\Hinf a(t')}
    \right], \quad \frac{k}{\Hinf a(t)} \ll 1.
\end{equation}
This is the form of the Green's function that one would use to
determine the behavior of $\Omega_\b$ at late times. It is also
interesting to consider the dynamics of modes which are in the
superhorizon regime at the \emph{beginning} of the de Sitter era:
$k/Ha(t') \ll 1$.  These are the modes which never actually enter
the Hubble horizon.  It is possible to explicitly evaluate the
residues of $\mathcal{G}_-$ in this case to obtain
\begin{equation}\label{eq:explicit G}
    G_-(t,t';0,z') \approx \Theta(\Delta t - \Delta z)
    \left\{ {A}_1 \left[ \frac{a(t)}
    {a(t')} e^{-\Hinf z'} \right]^{\epsilon/2}
    + {A}_2 \left[ \frac{a(t)}
    {a(t')} e^{-\Hinf z'} \right]^{3\epsilon/2-1/\Hinf r_\c}
    + \sum_{m=1}^\infty {B}_m \left[ \frac{k^2 e^{\Hinf z'}}{\Hinf^2 a(t) a(t') } \right]^{m}
     \right\},
\end{equation}
where $A_1$, $A_2$ and $B_m$ are functions of $\Hinf r_\c$. One can easily derive a similar expression for the advanced Green's function with coefficients $C_1$, $C_2$ and $D_m$ corresponding to the advanced modes in Table \ref{tab:resonances}.

The Green's function (\ref{eq:explicit G}) reveals how scale-dependence outside
the horizon can arise.  The $A_1$ and $A_2$ contributions to $G_-$ are independent of scale $k/Ha$.  This is consistent with
what we would expect in General Relativity for a mode that has $k/Ha
\ll 1$ at the beginning of a de Sitter era: the evolution of the
mode (as mediated by the Green's function) is expected to be
independent of $k$.  On the other hand, the contributions from the $B_m$ modes
carry an explicit $k$-dependence (this arises from the fact that
$J_{i\nu}$ and $J_{-i\nu}$ are linearly dependent when $i\nu$ is an
integer).
This $(k/Ha)^2$ suppression is also familiar from General Relativity and
represents the leading order correction to the gradient approximation.
However, on the normal branch, when combined with more slowly growing 
$A_1$ and $A_2$ modes, the net growth rate of the metric fluctuations for
superhorizon perturbations becomes $k$-dependent unlike in General Relativity.

We now turn our attention to finding a particular solution to the
inhomogeneous equation (\ref{eq:canonical system}) with the
$\Delta$ source. In principle, one could use the retarded Green's function
$G_-$ to find $\varphi^\particular$, but it is easier to just solve
for it directly in Eq.~(\ref{eq:canonical system}) by making a
physically-motivated ansatz.  In this calculation, we work in the $a
\rightarrow \infty$ limit and retain only leading order terms. We
make the following ``outgoing-wave'' ansatz for the bulk field:
\begin{equation}
    \varphi^\particular \propto e^{i\nu \Hinf (t-z)}.
\end{equation}
Substituting this into Eq.~(\ref{eq:canonical system}) and
(\ref{eq:Delta EOM}), we find that there are two values of $\nu$
consistent with $\Delta \ne 0$:
\begin{equation}
    \nu = 3i/2 \text{ or } 7i/2.
    \label{eq:particularscaling}
\end{equation}
The two distinct values of $\nu$ give rise to
solutions for $\Delta$ and $\Omega_\b^\particular$
\begin{equation}
    \Delta(t) = \Delta_0 + \Delta_1 (a/a_\star)^{-2}, \quad
    \Omega_\b^\particular \equiv a^{3/2} \varphi_\b^\particular =
    \frac{r_\c \kappa_4^2 \rho a^3}{\Hinf k^2}
    \begin{cases}
        \displaystyle \frac{\Delta_0}{3\Hinf r_\c-1} + \frac{1}{2}\frac{a_\star^2}{a^2}
        \frac{\Delta_1}
        {(5 \Hinf r_\c-1) }, & \epsilon = +1, \\ \displaystyle
        -2\Delta_0 + \frac{2}{3}\frac{a_\star^2}{a^2} \frac{\Delta_1}{(2 \Hinf r_\c-1)}, &
        \epsilon = -1,
    \end{cases}
\end{equation}
which confirms the scaling expectation [see Eq.~(\ref{eq:scalingparticular})].

Having now determined how to obtain both homogeneous $\varphi^\homo$
and particular solutions $\varphi^\particular$ of
Eq.~(\ref{eq:canonical system}) at late times, we can write down the
following formulae for various quantities of cosmological interest
\begin{equation}\label{eq:Omega decomposition}
    \Omega_\b \approx \sum_i \Omega_0^{(i)} \left( \frac{a}{a_\star}
    \right)^{p_i} + \Omega_\b^\particular, \quad \Phi
    \approx \sum_i \Phi^{(i)} + \Phi^\particular, \quad
    \Psi \approx \sum_i \Psi^{(i)} + \Psi^\particular.
\end{equation}
Here, $p_i$ is the temporal scaling index of the $i^\text{th}$
resonance of the \emph{retarded} Green's function (i.e.\ the
$p$-values for the $A_{1,2}$ and $B_m$ modes listed in Table
\ref{tab:resonances}), $\Omega_0^{(i)}$ are constants with dimension
$(\text{length})^2$ related to the amplitudes $(A_1,A_2,B_m)$ and recall 
$a_\star$ is the reference epoch at the beginning of the de Sitter phase [see Eq.~(\ref{eq:astar})].  The resonant contributions to the
metric potentials and comoving curvature perturbation are
\begin{subequations}\label{eq:resonance potentials}
\begin{align}
    \Phi^{(i)} & = \frac{\Omega^{(i)}_0 \Hinf^2
    (p_i-1)}{2a_\star(2\Hinf r_\c \epsilon - 1)} \left(\frac{a}{a_\star}\right)^{p_i-1} \left[ 1 + \mathcal{O}
    \left( \frac{k^2}{\Hinf^2 a^2} \right) \right], \\ \Psi^{(i)} &=
    (p_i-1)\Phi^{(i)} \left[ 1 + \mathcal{O}
    \left( \frac{k^2}{\Hinf^2 a^2} \right) \right].
\end{align}
\end{subequations}
For $p_i \ne 1$,
\begin{equation}
g_{\rm SH}^{(i)} = {p_i \over 2-p_i}
\end{equation}
for each mode on either branch.  This scaling is in fact a direct consequence of $\zeta'=0$ in
Eq.~(\ref{eq:zetascaling}) for $p_i \ne -2$.

On the self-accelerating branch with $\Omega_\Lambda=0$, the growth index
$p_i=2$  for both $A_1$ and
$A_2$. This large growth rate makes them dominates over $B$ and particular
modes.  In fact $g_{\rm SH} \rightarrow \infty$, which explains its divergent behavior in
Fig.~\ref{fig:QS first order}.
For the normal branch, the $A_1$ mode
has $p_i=1$ and thus no contribution to either $\Phi$ or $\Psi$.   The $k$-independent solution that dominates is then $A_2$ but its 
growth index  $p_i= -1/H r_\c$ is smaller
than the $B_1$ mode $p_i=1/2$.  Thus the $k$-dependent $B_1$ mode dominates
at late times for any finite $k$.

On the normal branch even the particular mode can be important.
The particular contribution to the various
quantities are given by
\begin{subequations}\label{eq:particular potentials}
\begin{align}
    \Phi^\particular & =
    \frac{\kappa_4^2 \rho a^2 \Hinf r_\c}{k^2}
    \begin{cases}
        \displaystyle +\frac{3}{2} \frac{\Delta_0}{3Hr_\c-1} , & \epsilon = +1, \\ \displaystyle
        -\frac{1}{3}\left( \frac{k}{Ha} \right)^2  \frac{\Delta_0}{2Hr_\c+1} +\left(\frac{a_\star}{a}\right)^2 \frac{2Hr_\c \Delta_1}{4H^2r_\c^2-1}, 
&
        \epsilon = -1,
    \end{cases} \\
    \Psi^\particular & =
    \frac{\kappa_4^2 \rho a^2 \Hinf r_\c}{k^2}
    \begin{cases}
        \displaystyle -\frac{3}{2} \frac{\Delta_0}{3Hr_\c-1}  , & \epsilon = +1, \\ \displaystyle
        +\frac{2}{3}\left( \frac{k}{Ha} \right)^2  \frac{\Delta_0}{2Hr_\c+1} -   \left(\frac{a_\star}{a}\right)^2 \frac{2(Hr_\c+1) \Delta_1}{4H^2r_
\c^2-1}
      , &
        \epsilon = -1.
    \end{cases}
    %\\ \zeta^\particular & = \cdots.
\end{align}
\end{subequations}
Note that for the normal branch the leading order $\Delta_0$ term
vanishes and contributions are suppressed by $(k/Ha)^2$.   Hence for
modes that are outside of the horizon at the de Sitter transition
$(k/H a_\star)^2 \ll 1$ we expect that the main contribution from
the particular mode comes from $\Delta_1$.  Under these assumptions the leading order scalings give
\begin{equation}
g_{\rm SH}^{(\rm p)}=
\begin{cases}
\displaystyle
0, & \epsilon=+1 \\
\displaystyle
 -{1 \over 2 Hr_\c +1},& \epsilon=-1 \,.
 \end{cases}
\end{equation}
For the normal branch, the growth index $p^{(\rm p)}=-2$ and hence can become the
leading order $k$-independent term if $-1/H r_\c < -2$.  However the value of $g_{\rm SH}$
remains the same as for the $A_2$ mode and is in fact the same as the matter dominated
scaling.  Therefore on the normal branch,
the leading order $k$-independent term can be modeled as
$g_{\rm SH}=-1/(2Hr_\c+1)$ regardless of which of those modes actually dominates the solution.  
This simplification was employed by \cite{Lombriser:2009xg} to model perturbation evolution to the present epoch
and obtain observational constraints on $H_0 r_\c$.

In Table \ref{tab:g}, we
list the values of $g_{\rm SH}$ associated with each of the modes appearing
in Eqs.\ (\ref{eq:resonance potentials}) and (\ref{eq:particular
potentials}), as well as the values of $g_{\rm SH}$ in matter domination from
\S\ref{sec:PL}.
These relations supplemented by the quasistatic analysis
complete the analytic description.

\begin{table*}
\newcolumntype{Q}{>{$\displaystyle}m{28mm}<{$}}
\newcolumntype{W}{>{$\displaystyle}m{12mm}<{$}}
\begin{tabular}{QQQQQQ}
\hline \text{branch} & \text{matter era} & A_1 & A_2 & B_1 &
\text{particular mode}
\\ \hline \hline%
\epsilon = +1 & +9/(8 H r_\c -1) & \infty & -\frac{3Hr_\c-1}{H r_\c-1} & 1/3 & 0 \\
\epsilon = -1 & -1/(2 H r_\c +1) & \text{n/a} & -1/(2 H r_\c +1) & 1/3 & -1/(2 H r_\c +1) \\
\hline
\end{tabular}
\caption{The value of the quantity $g$ in the superhorizon regime $k
/Ha \ll 1$ in either branch.}\label{tab:g}
\end{table*}

\begin{figure}\label{fig:sh}
\includegraphics[width=0.7\textwidth]{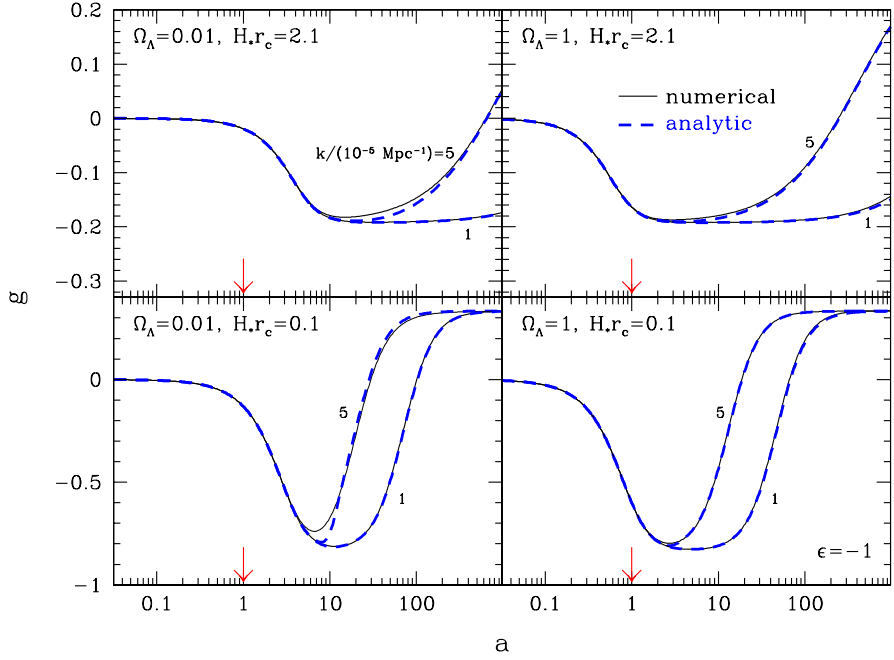}
	\caption{Metric ratio $g$ for purely superhorizon modes for normal branch ($\epsilon=-1$).  Analytic fit from Eq.~(\ref{eq:fullfit}) 
	(dashed lines) match the numerical CI calculation well for a wide range of parameters.  Note that even on superhorizon scales $g$ is scale dependent due to the relative importance of the $A$, $B$, and particular modes.    The current epoch is denoted by the arrow.}
	\label{fig:gsh}
\end{figure}

\section{Fitting Functions}
\label{sec:comparison}

Given the analytic description of perturbations in the matter and de Sitter epochs above and
below the horizon in \S \ref{sec:analytic},
we now devise global fitting functions for the perturbation evolution that bridge the transition and apply across
all scales and epochs.   We focus primarily on the normal branch but also test existing fits
for the self-accelerating branch that have been used to show its predictions 
are in conflict with CMB data  \cite{Hu:2007pj, Fang:2008kc}.

\subsection{Normal Branch}

As discussed in \S \ref{sec:perteqn}, once $g=(\Phi+\Psi)/(\Phi-\Psi)$ is determined,
solving for the metric itself is a simple matter of applying conservation of the comoving
curvature on large scales or the modified Poisson equation on small scales.
Moreover $g=0$ for General Relativity without anisotropic stress sources and an accurate
quantification of its value in DGP is important for observational test of gravity.

We begin by describing the superhorizon behavior.   As shown in the previous section,
there are several modes of importance during and after the transition to a 
de Sitter expansion.    Although the growth rate of all modes are independent of scale
above the horizon, their initial amplitudes are not.  In particular, after the de Sitter
transition at $a_\star$ where $H(a \rightarrow \infty) = H_\star = H(a_{\star})/\sqrt{2}$
 [see Eqs.~(\ref{eq:Hstar}) and (\ref{eq:astar})], there are scale-free modes and 
 modes that come from higher order terms in the gradient approximation that
 are suppressed by powers of $(k/H_\star a_\star)$.    The leading order
 mode $B_1$ has a faster growth rate and though initially suppressed will eventually dominate
 on all scales  [see Eq.~(\ref{eq:explicit G})].   $B_1$ requires
 $g=1/3$ whereas for
 the  scale free $A$ resonant and density-driven particular modes $g=-1/(2H r_\c+1)$.

\begin{figure}\label{fig:qs}
\includegraphics[width=0.7\textwidth]{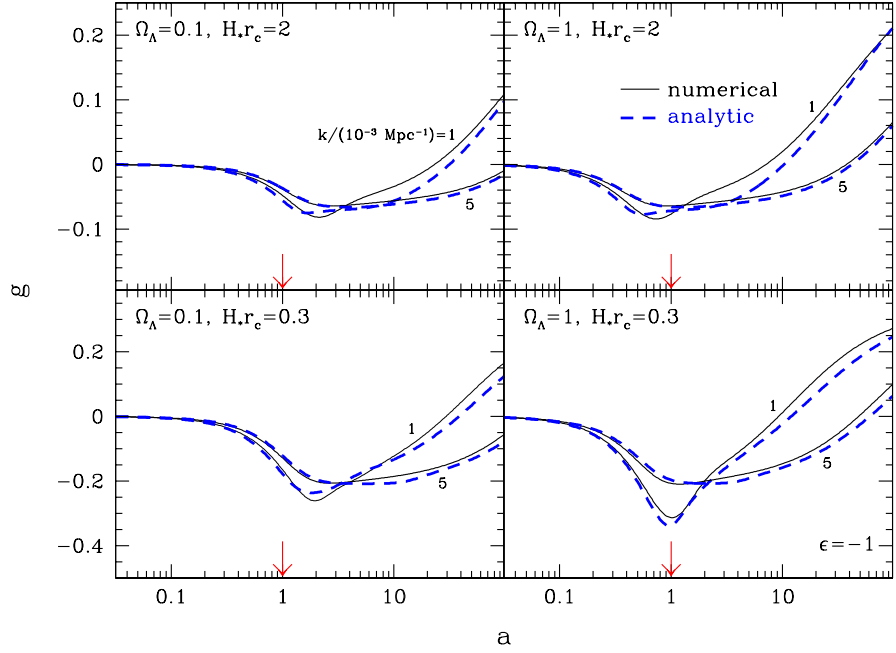}
	\caption{Metric ratio $g$ for modes that are currently horizon and sub-horizon scale for the normal branch ($\epsilon=-1$).  Analytic fit from Eq.~(\ref{eq:fullfit}) 
	(dashed lines) match the qualitative features of the numerical CI calculation across the full range of parameters.  In particular, the fit is designed to approach the correct behavior for
	$k/Ha \ll 1$, $k/Ha \gg1$ and $a/a_\star \gg 1$.  Note that currently subhorizon modes
	exit the horizon in the future de Sitter epoch.   The current epoch is denoted by the arrow. }
	\label{fig:gqs}
\end{figure}

 Given these
 relations and the growth rates of the three modes,
 we 
 seek to describe the superhorizon behavior as
\begin{equation}\label{eq:g SH fit}
g_\text{SH}(a,k) =  {(k/H_\star a_\star)^2  - [K_1 (a/a_\star)^{-1/2-1/H(a) r_\c} + K_2 (a/a_\star)^{-5/2}] 
\over {3(k/H_\star a_\star)^2 + [2 H(a) r_\c +1] [K_1 (a/a_\star)^{-1/2-1/H(a) r_\c} + K_2 (a/a_\star)^{-5/2}]}}.
\end{equation}
With 
\begin{subequations}\label{eq:K1 and K2}
\begin{eqnarray}
K_1 & = &  13.2 \left( { H_\star r_\c+1 \over H_\star  r_\c}\right)^2, \\
K_2 &=& 30.9 {( 2H_\star r_\c +1)^2 },
\end{eqnarray}
\end{subequations}
we are able to fit the results of the CI numerical solution very well from matter domination 
through de Sitter expansion for 240 models with $0.1 < H_\star r_\c < 10$, $0.01<\Omega_\Lambda<1$, and $k/H_\star a_\star \lesssim 1$.  Examples of the fit are shown in Fig.~\ref{fig:gsh}.

\begin{figure}
\includegraphics[width=0.6\textwidth]{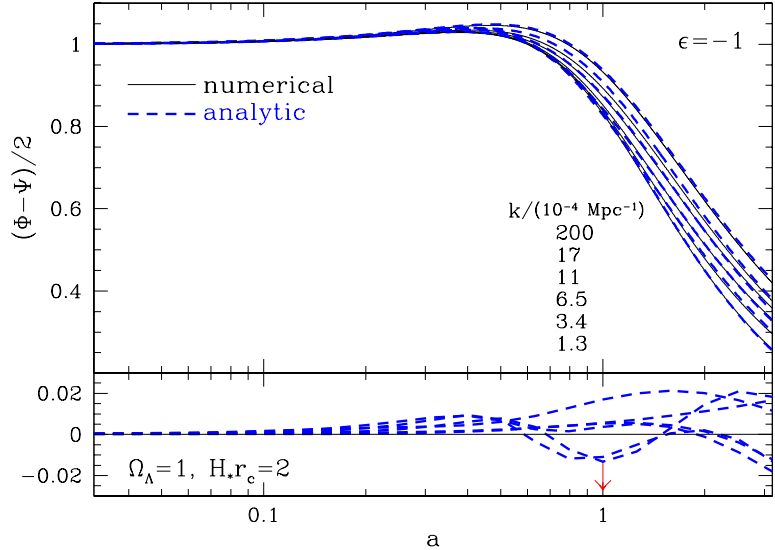}
	\caption{Observable lensing potential $(\Phi-\Psi)/2$ on normal branch ($\epsilon=-1$). Analytic fit to the metric ratio $g$ compared with the numerical CI calculation (top panel).   In spite of imperfections in the
	$g$-fit for currently horizon scale modes, fractional differences
	(bottom panel)  are
	$<2\%$ for $a<1$.     The current epoch is denoted by the arrow. }
	\label{fig:phim}
\end{figure}

The subhorizon form of $g$ approaches the quasistatic behavior at $k/Ha \gg 1$
\begin{equation}
g_{\rm QS}(a) = -{1 \over 3}{1 \over 1 - 2\epsilon Hr_\c(1+H'/3H)} .
\end{equation}
Note that in the de Sitter epoch, modes exit the horizon.  In choosing a functional form for
mediating the transition it is useful to note that
the first order predicted solution is $g=g_{\rm QS}( 1 + {\cal O}(Ha/k))$ from \S \ref{sec:QS}.   
Unfortunately it does not suffice to simply match a first order correction to the superhorizon modes at horizon crossing.  Instead we find that the following interpolation  
\begin{equation}
g(a,k) = { g_\text{SH} + g_\text{QS} K_3
\over 1 + K_3}
\label{eq:fullfit}
\end{equation}
with
\begin{equation}
K_3 = \left( 0.18 {k \over Ha} \right)^6 \left[ 1+ { ( 8.1 a/a_\star)^5 \over 3^5 + (k/H_\star a_\star)^5}(3 H_\star r_\c+1) \right]
\end{equation}
better describes the $k/Ha \sim 1$ horizon-crossing regime while providing a negligible mismatch
in the $k/Ha \gg 1$ limit.   Note that for $k/Ha \gg 1$ and $a/a_\star \gg 1$,
the first order correction of Eq.~(\ref{eq:qscorrection}) 
gives $(g-g_{\rm QS})/g_{\rm QS} = -2 ( H_\star a/k)$ whereas
the fit gives $(g-g_{\rm QS})/g_{\rm QS} = -1.7 (H_\star a /k) (1+H_\star r_\c)/(3+H_\star r_\c)$.
That the functional form in $k$ is correct and the $H_\star r_\c$ corrections are bounded
from $1/3$ to $1$ 
ensures that that we do not have a runaway mismatch for any set of parameters.

As shown in Fig.~\ref{fig:gqs}, the fit captures the qualitative features of the numerical
solution, including the asymptotic behavior at $a/a_\star \gg 1$.  Although imperfect,
the fit suffices for observational constraint purposes.   In Fig.~\ref{fig:phim}, we show the
performance of the fit for the metric combination involved in gravitational lensing
and gravitational redshifts $(\Phi-\Psi)/2$.   Here we have set the transition scale
between constant comoving curvature and Poisson-like behavior to $c_\Gamma=0.15$
\cite{Hu:2007pj}.  Note that fractional errors are below $2\%$ for $a<1$ for all
scales.

\begin{figure}
\includegraphics[width=0.6\textwidth]{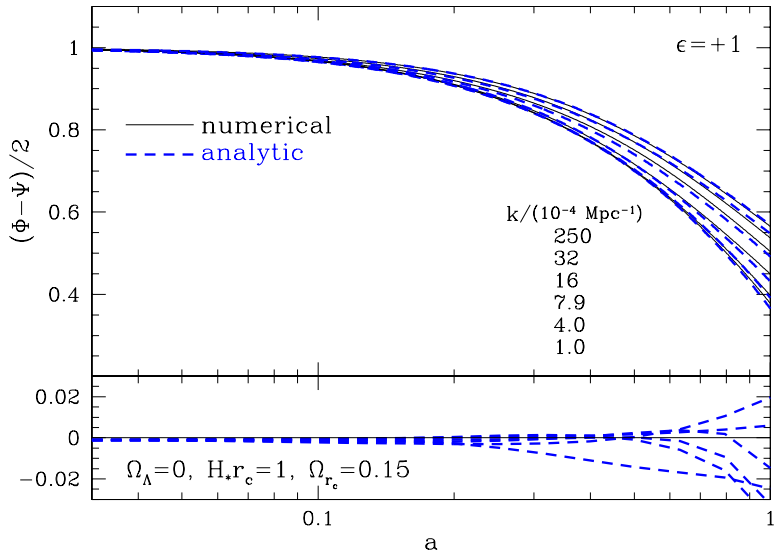}
	\caption{Observable lensing potential $(\Phi-\Psi)/2$ on self-accelerating branch
	($\epsilon=+1$).  Analytic fit to the metric ratio $g$ compared with the numerical CI calculation (top panel).   Fractional differences in the observable potential are
	$<2\%$ for $a<1$ (bottom panel).   }
	\label{fig:phimsa}
\end{figure}

\subsection{Self-Accelerating Branch}

Similar fitting functions for the self-accelerating branch were given in \cite{Hu:2007pj}
for $a<1$ and $\Omega_\Lambda=0$.  
For superhorizon scales
\begin{eqnarray}
g_{\rm SH}(a) & = & \frac{9}{8 H r_\c - 1} \left( 1 + \frac{0.51}{H r_\c - 1.08} \right). \nonumber
\label{eq:gSHsDGP}
\end{eqnarray}
and the interpolation factor in Eq.~(\ref{eq:fullfit}) is given by 
$K_3 = (0.14 k/Ha)^3$ with the transition scale set by $c_\Gamma=1$.
In Fig.~\ref{fig:phimsa}, we compare the predicted lensing potential to the numerical
integration up to the present epoch.   Given that we have shown that $g_{\rm SH}$ actually
diverges deep in the de Sitter epoch (see Tab.~\ref{tab:g} and Fig.~\ref{fig:QS first order}), it is not surprising the errors in the fit 
increase toward the current epoch.   Still those errors are $\lesssim 2\%$ at the time scales
relevant for observational constraints.   This should be compared with the much larger
change in the potential of a factor of $\sim 2$.

\section{Discussion}
\label{sec:discussion}

We have provided analytic solutions for the linear evolution of metric perturbations in the DGP 
modified gravity models in various regimes where scaling and Green's function techniques
are applicable.  We have elucidated the nature of the coordinate singularities and initial data in the bulk as well as  their effect on perturbation evolution on the brane.

Interestingly, even on superhorizon scales, the evolution of metric perturbations is no longer
necessarily scale free.  In the late-time de Sitter phase on the normal branch, several resonant modes are excited with different growth rates and amplitudes.
The epoch at which the fastest growing mode dominates depends on scale.  
On the other hand, for reasonable parameters where the de-Sitter phase has only recently been entered these complexities are mainly manifest in the future and do not affect observational tests \cite{Lombriser:2009xg}.

Based on these analytic solutions, we have devised convenient
fitting functions for the evolution that bridge the various spatial and temporal regimes for the normal branch.  
These forms are valid for essentially the whole range of parameter space
and have been tested against numerical integration of 240 models spanning
$0.1 < H_\star r_\c < 10$, $0.01<\Omega_\Lambda<1$ with multiple $k$-modes
for each.  For the self-accelerating branch we have verified the accuracy of
existing fitting formulae for the cases of interest \cite{Hu:2007pj}.

 Compared with a direct numerical integration of the bulk equations, 
these forms are  accurate at the percent level which is sufficient for current and upcoming  observational tests of the DGP 
scenario from the cosmic microwave background and weak gravitational lensing.

\vspace{1cm}
{\it Acknowledgments:}  We thank Y.S. Song for collaboration during the initial phases of this project.   SSS is supported by NSERC of Canada.
WH was supported by the KICP under NSF contract PHY-0114422, the DOE
contract DE-FG02-90ER-40560 and the Packard Foundation.
\bibliographystyle{arxiv_physrev}
\bibliography{ms}

\end{document}